\documentclass[journal,onecolumn]{IEEEtran}
\usepackage[utf8]{inputenc}
\usepackage[T1]{fontenc}
\usepackage{ragged2e}

\usepackage[cmex10]{amsmath}
\usepackage{amssymb}
\usepackage{array}
\usepackage{graphicx}
\usepackage{amssymb,amsmath}
\usepackage{cite}
\usepackage{multirow}
\usepackage{algorithm}
\usepackage{algpseudocode}
\usepackage{mathtools}
\usepackage{caption} 
\usepackage{ragged2e}
\newlength\ubwidth

\newcommand{\bor}[1]{{\color{red} #1}}
\newcommand{\irina}[1]{{\color{black} {#1}}}

\renewcommand{\vec}[1]{\ensuremath{\boldsymbol{#1}}}

\usepackage{tikz}
\usetikzlibrary{positioning}
\usetikzlibrary{babel}
\usepackage[a4paper]{geometry}
\pgfdeclarelayer{background}
\pgfdeclarelayer{foreground}
\pgfsetlayers{background,main,foreground}

\newtheorem{example}{Example}

\newtheorem{lemma}{Lemma}

\def\remark{
  \let\go\relax
  \ifvmode\vskip-\lastskip\fi
  \noindent{\it Remark\/.}%
  \enskip\relax\ignorespaces\go}
\newcommand{\bs}[1]{\ensuremath{\boldsymbol{#1}}}

\begin{document}

\title{ NB QC-LDPC  Coded QAM Signals with Optimized Mapping: Bounds and Simulation Results }

\author{
\IEEEauthorblockN{Irina E. Bocharova$^{1,2}$, 
  Boris D. Kudryashov$^{1,2}$,
  Evgenii P. Ovsyannikov$^3$, and \\
  {Vitaly Skachek$^2$}\\}
%  , and
%  T$\ddot{\text a}$hvend Uustalu$^2$ }
\vspace{1mm}
\IEEEauthorblockA{
	\small
	\begin{tabular}{c@{\hspace{.5cm}}c@{\hspace{.5cm}}c}
		               &   \\
		$^1$University of Information  & $^3$State University of  & $^2$University of Tartu, Estonia \\
		 Technologies, Mechanics   & Aerospace Instrumentation&Email: \{irinaboc, boriskud\}@ut.ee   \\
	and Optics	&  St. Petersburg,190000,  Russia & \{vitaly.skachek\}@ut.ee\\
	St. Petersburg, 197101, Russia     	  &  Email: {eovs@mail.ru}&\\
		\end{tabular}
	\vspace{-4mm}
}
}

\maketitle
\renewcommand{\thefootnote}{\fnsymbol{footnote}}
\footnotetext{ Parts of this work were presented  at the IEEE ISIT 2021 \cite{bocharova2021euclidean} and ITW 2021 \cite{bocharova2021random}. 

The work of I.E. Bocharova, B.D. Kudryashov and V. Skachek is supported in part by the grant PRG49 from the Estonian Research Council. The work of V. Skachek is also supported in part by the ERDF via CoE project EXCITE. The work of I.E. Bocharova, B.D. Kudryashov and E.P. Ovsyannikov is also supported in part by the Ministry of Science and
Higher Education of Russian Federation, project no. 2019-0898.}
\renewcommand{\thefootnote}{\arabic{footnote}}
\begin{abstract}

This paper studies specific properties of nonbinary low-density parity-check  (NB LDPC) codes when used in coded 
modulation systems. The paper is focused on the practically important NB LDPC codes over extensions of the Galois field GF$(2^m)$ with $m\le 6$ used with QAM signaling.   Performance of NB QC LDPC coded transmission strongly depends 
on mapping of nonbinary symbols to signal constellation points. We obtain a random coding bound on the maximum-likelihood decoding error probability for  an ensemble of random irregular NB LDPC codes used with QAM signaling for specific symbol-to-signal point mappings. This bound is based on the ensemble average squared Euclidean distance spectra derived for these mappings.
The simulation results for the belief-propagation decoding in the coded modulation schemes with the NB quasi-cyclic  (QC)-LDPC codes under different mappings are given. Comparisons with the optimized binary QC-LDPC codes in the WiFi and 5G standards, as well as with the new bound, are performed.

%We study two approaches to probabilistic shaping  of  $M$-QAM modulated signals used in conjunction with binary images of nonbinary (NB)  quasi-cyclic  (QC) LDPC codes over $GF(2^m)$.  We start with  optimization  code symbol-to-QAM signal mapping  for the unshaped QAM signals. Then two multi-dimensional modifications of probabilistic shaping  based on matching between  modulation and field order, that is, $M$ and $m$, are introduced.   Random coding bounds  on the error performance  for ensembles of  irregular NB LDPC codes of finite lengths  over $GF(2^m)$ used in the AWGN channel with unshaped and shaped QAM signaling are derived. The simulated FER performance of the sum-product BP decoding of  irregular  NB QC-LDPC block codes  with shaped and unshaped QAM signaling  are presented and compared with the derived finite-length random coding bounds as well as with  the same performance of the optimized  binary QC-LDPC block code in the WiFi  and 5G  standards.   

\end{abstract}

\section{Introduction}
%%%%%%%%%%%%%%%%%%%%%%%%%%%
Nonbinary (NB) LDPC block codes over extensions of the binary Galois field were introduced  in \cite{mackay}. Since that time, the term  {\em NB LDPC codes} is used for binary images of NB LDPC codes  over extensions of the binary  field \irina{\cite{poulliat2006using}}, unlike the NB LDPC codes over  arbitrary fields in \cite{gallager}. In this paper, with a slight abuse of notations, we use the terms binary images of NB LDPC codes and NB LDPC codes interchangeably.  

\subsection{NB LDPC codes  with small alphabets }

The advantage of NB LDPC codes compared to their binary counterparts when used with QAM signaling in the AWGN channel  was demonstrated in a number of papers (see, for example, \cite{declercq2004regular}, \cite{pfletschinger2009performance}, \cite{mostari2015simplified}  and references therein).
In optimization of  NB LDPC codes with both BPSK and  QAM signaling,  the most attention was paid to  so-called `ultra-sparse' regular codes, that is, the regular codes  with only two nonzero elements in each column of their parity-check matrices. Mainly the Galois field extensions GF$(2^m)$ with $m\ge 6$ were considered since it was shown by simulations of NB LDPC codes of short and moderate lengths that increasing $m$ improves the performance of iterative decoding. 
The belief-propagation (BP) decoding thresholds for ultra-sparse regular NB LDPC codes used in the AWGN channel as a function of $m$ were presented in \cite{poulliat2006using} and \cite{poulliat2008design}.
It was shown  that BP decoding thresholds  \irina{for the AWGN channel} strongly depend on the field size and are much smaller for $m\ge 6$ than those for smaller $m$, \irina{although in the general, behavior of the thresholds  may not  be a monotonic function.} However, increasing $m$ leads to the higher computational complexity of BP decoding. This draws attention to looking at NB LDPC codes over smaller fields to find a trade-off between performance and complexity.

NB LDPC codes with QAM signaling for small field sizes, that is, for $m=2,...,5$ were considered in a limited number of papers. It was discovered through the analysis by density evolution technique  as well as simulations  in \cite{li2009density} that for small $m < 6$, the average column weight of  NB LDPC  code base matrix should be 
around the interval [2.2,~2.4]. In the same paper, BP decoding thresholds for rate 1/2 `semi-regular' NB LDPC codes over GF$(2^m)$, $m=2,...,6$ were derived. The authors concluded that if column weight is larger than 3,  the thresholds  for small field sizes are better than for large ones, while for column weight smaller than 2.5, performance improves with  increasing $m$. In \cite{arabaci2009high}, experimental results for high-rate regular NB QC-LDPC codes over GF$(2^2)$, GF$(2^3)$, and \irina{GF$(2^4)$} used for optical communications were presented. High-rate irregular NB LDPC codes were studied in \cite{rezqui}. It was noticed in this paper that high-rate NB LDPC codes with three nonzero elements in each column of their parity-check matrix are superior to ultra-sparse NB LDPC codes if the alphabet size is small.   

In \cite{bocharova2020optimization}, an approach for optimization of NB QC LDPC block 
codes over GF$(2^m)$ for $m\le 6$ was suggested. It is based on applying the simulated annealing technique \cite{delahaye2019simulated}  to constructing the code base matrix followed by optimization of the degree matrix  performed by the algorithm  in \cite{Boch2016}. In the same paper, an ensemble of irregular NB LDPC codes determined by the parity-check matrix with column weights  two and three was introduced and analyzed. A finite-length random coding bound on the maximum-likelihood (ML) decoding performance for this ensemble of codes used with BPSK signaling in the AWGN channel was derived.

\irina{
The improvement in the performance of NB LDPC codes over their binary counterparts
comes at the cost of higher decoding complexity.
In the general case, the generalized belief propagation (BP) decoding complexity per bit for a binary image of the NB LDPC code is proportional to $q^2$, where $q=2^m$ . However, implementation of the decoder based on the fast Hadamard transform  (FHT) has complexity proportional to $q \log_2 q$ (see, for example, \cite{declercq2007decoding}). Moreover, there exist simplified decoding techniques such as, for example, extended min-sum (EMS) algorithm proposed in \cite{declercq2005extended}, \cite{declercq2007decoding}, 
or Min-Max decoding in \cite{savin2008min},  \cite{zhang2010partial},
 which allow to significantly reduce the decoding complexity  at the cost of relatively  small losses in the decoding performance ($\approx 0.2$ dB). According to the published results regarding  VLSI implementation complexity \cite{lacruz2014simplified},  
simplified decoding of NB LDPC codes over GF($2^4$) is a few times more complex than the decoding of binary 
codes of the same rate and binary length with  the same throughput of the decoder. 
}
\subsection{Matching  NB LDPC codes with  QAM modulation }

 It was shown (see, for example, \cite{declercq2004regular}) that compared to the binary LDPC codes used with QAM signaling, the performance of NB LDPC  codes used with QAM signaling is much more influenced by the choice of code symbol-to-QAM signal mapping. This, typically, restricts the number of admissible pairs $(m, M)$ used both in practical schemes and theoretical studies.  In particular, in \cite{declercq2004regular}, regular NB LDPC codes over GF$(2^m)$ with  $M^2$-QAM signaling, where  either $\log_{2}M=m$  or $\log_{2}M$ is  a divisor of $m$, were considered. 
 
Since  QAM signals are represented by two orthogonal PAM-signal components, the achievable transmission rate for
QAM signaling in the AWGN channel is precisely  two times larger than for PAM signaling with the same signal-to-noise ratio per signal component.  %In the sequel, we study only PAM signals. 
In the survey below, we do not distinguish between research results for 
%We also refer to 
QAM and PAM 
%interchangeably 
meaning that results obtained for  $M$-PAM signaling can be easily reformulated for $M^2$-QAM signaling.      

There are two main techniques for matching  outputs of a binary LDPC encoder with inputs of the QAM modulator: bit-interleaved coded modulation (BICM) \cite{caire1998bit} and multi-level coding (MLC)~\cite{imai1977new}. When using BICM, the encoded bits are interleaved and mapped, typically according to a Gray mapping,  to  signal constellation points.  The MLC mapping can be considered as a generalization of trellis-coded modulation (TCM) introduced in \cite{ungerboeck1982channel}.  When applying this technique, the input bits are split into groups, and bits inside each group can be encoded by different codes or left uncoded.  Then the encoded bits  of the same group are mapped into appropriate signal constellation points.  

In order to match the NB LDPC code  with QAM signals besides  BICM,  symbol interleaved modulation (SICM), as in \cite{declercq2004regular}, \cite{pfletschinger2009performance } and \cite{rezqui}, is used. In this case, symbols of  an NB LDPC code are  interleaved and then  mapped to  modulation signals.
 
This paper is organized  as follows. In Section \ref{prelim}, necessary definitions are given. QAM modulation and demodulation schemes in combination with BP decoding for irregular NB LDPC codes over small alphabets are studied in  Section {\ref{Demodulation}}.   In Section \ref{AlmostRegular},  we  consider an ensemble of random irregular NB LDPC codes and a random binary image of this ensemble used with the QAM signaling and different mappings in the AWGN channel. The average squared Euclidean distance spectra (SEDS) for these two ensembles are derived and discussed in the same section.  A finite-length random coding  bound  on ML decoding  error probability for the ensemble of irregular NB LDPC codes  based on its SEDS is obtained.  
Simulation results of the FER performance of the BP decoding for the NB QC-LDPC codes, optimized as in  \cite{bocharova2020optimization}, are presented and compared with the theoretical bound in Section \ref{Simul}.
The paper is concluded by a short discussion.  For completeness,  a known  bound  on the error probability of ML decoding is presented in Appendix.  

The main contributions of the paper are:
\begin{itemize}
\item{The new modulation-demodulation schemes suitable for any pairs of parameters $(m,M)$}
\item{Average squared Euclidean distance spectra for the ensemble of irregular NB LDPC codes used with BICM and SICM mappings}
\item{A finite-length random coding bound on the error probability of ML decoding for the ensemble of irregular NB LDPC codes used with QAM signaling and different mappings in the AWGN channel}

\end{itemize}

\section{Preliminaries \label{prelim}}

A  rate $R=b/c$  NB QC-LDPC code over GF($2^m$)  is defined by its polynomial parity-check matrix of size $(c-b)\times c$ 
\[
H(D) = \{ h_{ij}(D) \} \;,
\] 
where  $h_{ij}(D)$ are polynomials of formal variable $D$ with coefficients from GF($2^m$). 
In the sequel,  $h_{ij}(D)$ are either zeros or monomials  and 
\[
H(D) = \{ \alpha_{ij}D^{w_{ij}} \} \;, w_{ij}\in \{0,1,...,\nu\}\;,  \alpha_{ij} \in {\rm GF}(2^m),i=1,...,c-b,j=1,...,c,
\] 
where  $\nu$ denote  the maximal degree of a monomial.  
The corresponding   $q$-ary parity-check matrix, $q=2^m$ of the $(Lc,Lb)$ NB QC-LDPC block code is obtained by replacing $D^{w_{ij}}$,  by the $w_{ij}$-th power of a circulant permutation matrix of order $L$. The parameter $L$ is called 
{\em lifting factor}. 
The parity-check matrix in binary form \irina{which determines {\em binary image} of the NB LDPC code}   is obtained  by replacing  non-zero elements of the $q$-ary, $q=2^m$ parity-check matrix by  binary $m\times m$ matrices, which are companion matrices of the corresponding field 
elements~\cite{macwilliams1977theory}.

Let $\vec \alpha_i = (\alpha_{i1},\alpha_{i2},...,\alpha_{iw_i})$ be a vector consisting of nonzero elements of 
$i$th row of $H(D)$ and $w_i$ be the number of nonzero elements of this row. 
After replacing these nonzero elements with their binary $m\times m$ companion matrices, we obtain an 
$m\times mw_i$
parity-check matrix of a linear code which  we call the  $i$-th {\em constituent} code of the NB LDPC code.

To facilitate the low encoding complexity, we consider 
parity-check matrices having the form (see, for example, \cite{Boch2016})
 \begin{equation} \label{bidiag}
H(D)=
\begin{pmatrix}
 H_{\rm inf}(D) & \vec h_0(D)& H_{\rm bd}(D) 
\end{pmatrix},
\end{equation}
where  $H_{\rm bd}(D)$ is a bidiagonal matrix of size $(c-b)\times (c-b-1)$,
$\vec h_0(D)$ is a column with two % at most 3 
nonzero elements, %, two of which are identical monomials,
and $H_{\rm inf}(D)$ can be any monomial submatrix of \irina{size $(c-b)\times b$}.
This submatrix corresponds to the  information part of a codeword.
  
Binary matrix $B=\{b_{ij}\}$ of the same size as $H(D)$ is called {\em base} matrix for $H(D)$ if 
$ b_{ij}=1$ iff $h_{ij}(D)\neq 0$.

In the search  for optimized parity-check matrices, we represent  $H(D)$ in the form of two matrices: degree matrix $H_w=\{ w_{ij} \}$
and matrix of field coefficients $H_c=\{ \alpha_{ij} \}$ which we obtain by labeling nonzero elements of  $B$  by monomial degrees and nonzero field elements, respectively. In these matrices  
{only elements for which elements}  
of base matrix $b_{ij}=1$ are meaningful. 
For that reason, in $H_w$ and $H_c$  we write ``$-1$'' in positions corresponding 
to zero elements of $B$.   

\irina{For example, the rate $R=1/4$ NB QC-LDPC  code over GF$(2^4)$ determined by \[H(D)=\left(\begin{array}{l l l l}
D^0& D^0  &        D^0  &    D^0\\
D^0 &\alpha D^3 &   0 &\alpha^{3}D^7\\
D^0 &\alpha^{2}D^9 &    \alpha^{5}D^2 &\alpha^{11} D
\end{array}\right)
\]
has the following matrices $H_{w}$ and $H_c$ 
\[H_{w}=\left(\begin{array}{r r r r }
0& 0  &       0  &    0\\
0 &3 &   -1 &7\\
0 &9 &   2 &1
\end{array}\right),
\]
\[H_{c}=\left(\begin{array}{r r r r }
1& 1  &       1  &   1\\
1 &\alpha &   -1 &\alpha^3\\
1 &\alpha^{2} &    \alpha^{5} &\alpha^{11}
\end{array}\right),
\]
where $\alpha$ is a primitive element of GF$(2^4)$.
The corresponding matrix  $B$ is
\[B=\left(\begin{array}{r r r r }
1& 1  &      1  &    1\\
1 &1 &   0 &1\\
1 &1 &  1 &1
\end{array}\right).
\]
}

Notations used throughout the paper are summarized in Table \ref{Notations}.

\begin{table}
%\bor{
\caption{Notations \label{Notations}}
\begin{tabular}{l|l}
\hline
Notation & Comment \\
\hline
$b$  & Number of rows in the base matrix\\
$c-b$  & Number of columns in the base matrix\\
$B=\{b_{ij}\},i=1,...,c-b,j=1,...,c$ & Base matrix \\
$D$ & Formal variable \\
$H_w=\{w_{i,j}\},i=1,...,c-b,j=1,...,c$ & Degree matrix, $w_{ij}=-1$ corresponds to zero entry in $B$\\
$H_c=\{\alpha_{i,j}\},i=1,...,c-b,j=1,...,c$ & Matrix of coefficients, $\alpha_{ij}=-1$ corresponds  \irina{to} zero entry in $B$\\
$q=2^m$ & Size of extension of the Galois field  \\
$n,k, r=n-k$  & Codelength, number of information symbols and number of redundant bits \\
$N= \frac{n}{m}$ & Codelength in symbols of GF($q$) \\
$M=2^p$  & PAM modulation order. It corresponds to $M^2$-QAM \\
$N_{p}=\frac{n}{p}$ & Length of codeword in PAM signals\\
$L=\frac{n}{c}$ or $L=\frac{n}{K}=\frac{r}{J}$ & Lifting factor of QC-LDPC code or strip size for  Gallager's  ensemble\\
$(J,K)$               & Column and row weight for regular LDPC code  \\
$\mathcal L(\cdot)$  & Log-likelihood ratios for symbols and sequences \\  
AE & Signal alphabet extension \\
ASCM& Amplitude-sign coded modulation \\
BICM& Bit-interleaved coded modulation \\
BPCM& Bit-plane coded modulation\\
CGF &Combinatorial generating function \\
MGF& Moment generating function\\
SICM& Signal-interleaved coded modulation \\
SEDS& Squared Euclidean distance spectrum\\
SED& Squared Euclidean distance  \\ \hline 
 \end{tabular}
 %}
\end{table}

\section{Mappings of NB LDPC code symbols to QAM signals \label{Demodulation}}

In order to provide for better matching of the PAM signals with the NB LDPC codes, one should take into account the fact that the sign and the amplitude bits of the PAM signals have different significance. This property makes the error-correcting capability of the NB LDPC codes sensitive to the mapping of the code symbols onto the PAM signals. Next, we consider four modulation schemes matched with NB LDPC coding. 
%As it is mentioned above, when constructing NB LDPC codes over relatively small fields GF$(2^m)$, $m<6$ the  analysis by density evolution technique, as well as simulations, show that the average column weight of base matrix should be 
%around the interval [2.2,~2.4]. For this reason in this section we consider NB LDPC codes whose parity-check matrices have the aforementioned bi-diagonal structure and columns with  two and three nonzero elements only. 

\subsection{PAM mapper-modulators}
\begin{figure}
\begin{center}
\includegraphics[width=130mm]{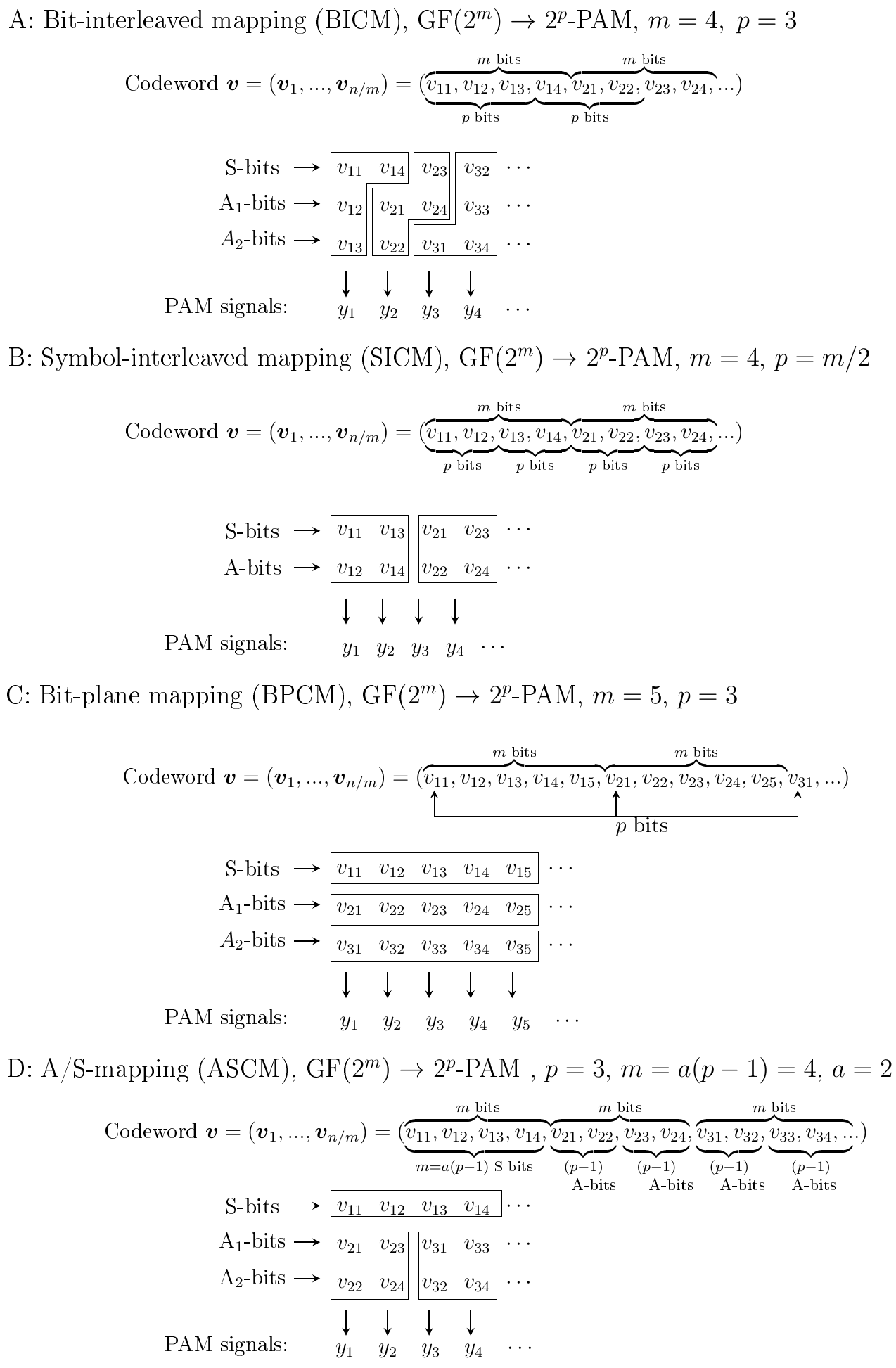}   
\caption{\label{Mappings} Modulation techniques for matching outputs of NB LDPC codes over GF$(2^m)$ 
and $2^p$-PAM signals, $p=\log_2 {M}$   }
\end{center}
\end{figure}

%There are two main properties of PAM and shaped PAM signals which should be taken into account in order to provide  better matching with NB  LDPC codes. Sign and amplitude bits of PAM signals have different significance. Sign bits being most significant bits, are more reliable than amplitude bits. 
%The second property  is related to shaping which is applied only to  amplitude bits. These properties  make error correcting capability of NB LDPC codes sensitive to mapping code symbols to PAM signals.
Let $n$ be the codelength in bits. Suppose that both $m$ and $p=\log_2M$ are divisors of $n$, that is, $N=n/m$ and $N_p=n/p$ are integers.  Four modulation schemes matched with non-binary coding
are presented in  Fig. \ref{Mappings}.  The modulator A performs a bit-interleaved mapping (BICM), that is, codeword bits $v_{11},v_{12},...$ corresponding to $N=n/m$ $2^m$-ary symbols 
are split into groups of $p<m$ bits,  where the first bit of each group \irina{($S$-bit)} is mapped  onto the sign of  the $2^p$-PAM signal and the other bits \irina{($A_1$,$A_2$,$\dots$,$A_{p-1}$)}  determine  the amplitude of the corresponding $2^p$-PAM signal. 
Modulator A can be used for any pair $(m,p)$.  If this mapping is used, typically, code symbols are not equally reliable due to an arbitrary mapping of their bits onto signs and amplitudes of the PAM signals.

Modulator B implements symbol-interleaved mapping (SICM). It is applicable only if $p$ is either equal to $m$ or is a 
divisor of $m$.  We consider the latter case. Then each group of $m$ bis corresponding to the $2^m$-ary symbol 
is split into a few groups of $p$ bits. The first bit in each group determines the sign of  the $2^p$-PAM signal, and 
the other bits of each group are mapped onto the PAM signal amplitude. 
Thus, a certain fixed number of sign bits are combined with a certain number of 
amplitude bits. This makes all code symbols equally reliable. 

\irina{The next two new mappings do not require $p$ to be a divisor of $m$ and can be applied to a large set of parameters $m$ and $p$. We show both by comparing theoretical ML decoding error probability bounds and by simulating  BP decoding  that, unlike the BICM, these  mappings used with {\em {modified}} demodulators  contribute to  better decoding performance.}
 
Now assume that $p>m$, $m$ is not required to be a divisor of $p$.   
The modulator C performs ``{\em a bit-plane mapping}'' (BPCM), which assumes that  the first, the $(p+1)$-th,  the $(2p+1)$-th, $\dots$ group of $m$ code bits are mapped onto the signs of the PAM signals, and other groups of $m$ code bits are mapped onto the first,  the second, 
$\dots$, the $(p-1)$-th amplitude bits of the PAM signals.  The shortcoming of this mapping is that symbols consisting of sign (most significant) bits are more reliable than the other symbols consisting of amplitude (less significant) bits.  
However, if BPCM is used, the reliability distribution of the code symbols can be controlled both in the demodulator and at the code design level.

Mapping  performed in the modulator D we call ``{\em amplitude/sign mapping}''  (ASCM). It is assumed that $p<m$ and $p-1$ is a divisor of $m$, that is, $m=a(p-1)$, $a$ is an integer.   The sign bits of the $m$ PAM signals are formed as in BPCM.  Then the first, the second,..., the$(p-1)$-th amplitude bits of  the $a=m/(p-1)$ PAM signals are formed. Similarly to BPCM, ASCM provides a possibility to control the code symbol reliabilities.

\begin{table} 
\centering
\caption{Gray mapping amplitude to amplitude bits \label{Gray}}
\begin{tabular}{|c|c|c|c|c|c|c|c|c|}
\hline
Modulation & \multicolumn{8}{c|}{Binary representation \irina{$\vec v$} for amplitudes \irina{$A(\vec v)$} of PAM signal points} \\ \cline{2-9} 
order         & 1 & 3 & 5 & 7& 9 & 11 & 13& 15  \\ \hline
4-PAM       & 1 & 0 &-- &-- &-- &--   & -- & -- \\ \hline 
8-PAM       & 10 & 11 & 01 &00 &-- &--   & -- & -- \\ \hline 
16-PAM     & 110 &111&101 &100 &000 &001& 011 & 010 \\ \hline 
\end{tabular}
\end{table}

In  the $2^p$-PAM modulator, first,  one of the
four  mappers (A--D in Fig. \ref{Mappings}) is used to form $p$ binary sequences: one sign bit sequence $\bs s=(s_1,s_2,...,s_{N_p})$ and $p-1$ amplitude
bit sequences $\bs a_i=(a_{i1},a_{i2},...,a_{iN_p})$, $i=1,...,p-1$. 
Then, the PAM signal  $y_t$ is computed as
\[
y_t=(2s_t-1) A (a_{1t},...,a_{(p-1)t}), \quad t=1,2,...,N_p,
\]   
where $A( \cdot) $ denotes the amplitudes of  the signal points obtained according to the Gray mapping. An example of the Gray mapping used for 4-PAM and 8-PAM modulation is given in Table \ref{Gray}. 

The binary image of a $2^m$-ary codeword $\bs v=(\bs v_1, \bs v_2,\dots,\bs v_N)$, $\bs v_i=(v_{i1},\dots, v_{im})$ is mapped onto the signal sequence $\bs y=(y_1,y_2,\dots,y_{N_p})$, which is transmitted over the AWGN channel.  The  received sequence is
\[
r_t=y_t+n_t\;, \quad t=1,2,...,N_p,
\]
where the noise samples $n_t$ are independent Gaussian variables with zero mean and variation
$\sigma^2=N_0/2$.  
 
 \subsection{Demodulator-demappers} 
 \irina{In this subsection,  we show that for different mappings dependencies between NB code symbols and the corresponding PAM signals can be taken into account when computing symbol LLRs. A decoding performance gain can be obtained by using more informative outputs of the demodulator. 
 
 First, we revisit the demodulator for the BICM mapping and then show how   the reliability of the $2^m$-ary code symbol can be computed when the mapping onto a few PAM signals is applied.}   
In the demodulator, first, LLRs of signal points are computed. 
We assume that signal points are equiprobable, then the a \irina{posteriori} probabilities
of signal points $\xi \in\{-2^p+1$,$-2^p+3$,$\dots$, $-1,1,\dots$,$2^p-3$, $2^p-1\}$ given the channel output $\bs r=(r_1,...,r_{N_p})$ are
\[
P(y_t=\xi|r_t)=K(r_t)\exp \left\{-\frac{(r_t-\xi)^2}{N_0}\right\}\;,
\]
where $K(r_t)$ are probability normalizing coefficients providing $\sum_\xi P(\xi|r_t)=1$.
Next, a posteriory probabilities  of bits in  the binary representation of signal points are computed 
by summing up probabilities of signal points with zero and one in the corresponding positions. 
%For example, for 8-PAM, according to Table \ref{Gray} 
%\begin{eqnarray}
%L(s|r_t)&=&\log \frac{P(1|r_t)+P(3|r_t)+P(5|r_t)+P(7|r_t)}
%                     {P(-1|r_t)+P(-3|r_t)+P(-5|r_t)+P(-7|r_t)} \label{LLRsbit} \\
%L(a_1|r_t)&=&\log \frac{P(-3|r_t)+P(-1|r_t)+P(1|r_t)+P(3|r_t)}
                    % {P(-7|r_t)+P(-5|r_t)+P(5|r_t)+P(7|r_t)}\label{LLRa1bit}\\
%L(a_2|r_t)&=&\log \frac{P(-5|r_t)+P(-3|r_t)+P(3|r_t)+P(5|r_t)}
 %                    {P(-7|r_t)+P(-1|r_t)+P(1|r_t)+P(7|r_t)} \label{LLRa2bit}
%\end{eqnarray}
According to Table \ref{Gray}, the log-likelihood ratios (LLRs) $\mathcal L(\cdot)$ of the sign and amplitude bits in the binary representation of the signal points $\xi$ are computed as follows 
\begin{eqnarray}
\mathcal L(s|r_t)&=&\log \frac{P_{\xi \in \xi_{\rm s,1}}(\xi|r_t)}{P_{\xi \in \xi_{\rm s,0}}(\xi|r_t)}\label{LLRsbit}\\
\mathcal L(a_i|r_t)&=&\log \frac{P_{\xi \in \xi_{{\rm a}_{i},1}}(\xi|r_t)}{P_{\xi \in \xi_{{\rm a}_{i},0}}(\xi|r_t)} \label{LLRa1bit},
\end{eqnarray}
where $i=1,...,p-1$, 
\[
\xi_{\rm s,0}=\{-2^p+1,-2^p+3,\dots,-3,-1\},
\]
\[
\xi_{\rm s,1}=\{1,3,\dots ,2^p-1\},
\]  
\[
\xi_{{\rm a}_{1},0}=\{-2^{p}+1,-2^{p}+3,\dots,-2^{p-1}-1, 2^{p-1}+1,2^{p-1}+3,\dots, 2^{p}-1 \}.
\]
\[
\xi_{{\rm a}_{1},1}=\{ -2^{p-1}+1, -2^{p-1}+3, \dots,-1,1,3,\dots, 2^{p-1}-1 \}.
\]

For example, for 4-PAM used with the Gray mapping as in Table \ref{Gray}, $\xi_{\rm s,0}=\{-3,-1\}$, $\xi_{\rm s,1}=\{1,3\}$, $\xi_{\rm a_{1},0}=\{-3,3\}$, $\xi_{\rm a_{1},1}=\{-1,1\}$.

 Signal sets $\xi_{{\rm a}_{i},0}$ and  $\xi_{{\rm a}_{i},1}$ for $i\ge 2$ depend on the modulation index. For example, for 8-PAM $\xi_{{\rm a}_{2},0}=\{-5,-3,3,5\}$ and  $\xi_{{\rm a}_{2},1}=\{-7,-1,1,7\}$. According to Table \ref{Gray}, we have
\begin{eqnarray}
\irina{\mathcal L}(s|r_t)&=&\log \frac{P(1|r_t)+P(3|r_t)+P(5|r_t)+P(7|r_t)}
                     {P(-1|r_t)+P(-3|r_t)+P(-5|r_t)+P(-7|r_t)} \label{LLRsbitT} \\
\irina{\mathcal L}(a_1|r_t)&=&\log \frac{P(-3|r_t)+P(-1|r_t)+P(1|r_t)+P(3|r_t)}
                    {P(-7|r_t)+P(-5|r_t)+P(5|r_t)+P(7|r_t)}\nonumber\\
\irina{\mathcal L}(a_2|r_t)&=&\log \frac{P(-5|r_t)+P(-3|r_t)+P(3|r_t)+P(5|r_t)}
                    {P(-7|r_t)+P(-1|r_t)+P(1|r_t)+P(7|r_t)} \nonumber.
\end{eqnarray} 
 
\irina{Computation of} the  LLRs for the NB  code symbols $\bs v$ depends on the mapping used. In the case of BICM, one, typically, ignores the symbol bits dependencies, and the LLRs of code symbols $\bs v$ are computed as:
\begin{equation}
\mathcal L(\bs v=(v_1,...,v_m)|\bs r)=\sum_{i=1}^m\mathcal L(v_i|r_i) \label{BICM}.
 \end{equation} 
When BICM is applied, if the bits are randomly permuted before mapping, expression (\ref{BICM}) gives the correct LLR values. 

If the bits of the $2^m$-ary symbol are mapped onto a few PAM signals, then the amplitude bit dependencies can be taken into account. For BPCM, we compute the symbol LLRs as:
\begin{equation}
\mathcal L(\bs v=(v_1,...,v_m)|\bs r)=\sum_{i=1}^{m}
\mathcal L(v_i|\bs r) \;. \label{LLRsymbol} 
\end{equation}  
In terms of the sign and amplitude bits of the PAM signals we have 
\[\mathcal L(\bs v_j|\bs r)=\sum_{i=1}^m\mathcal L(v_{ji}|\bs r) \mbox{, }
v_{ji}=\left \{
\begin{array}{cc}
s_{(j-1)m+i}\mbox{, } &j=1,p+1,...,N-p+1\\
a_{\ell,(j-1-\ell)m+i}\mbox{, } & j=\ell+1,\ell+p+1,...,N-p+\ell+1
\end{array},
\right. 
\]
$j=1,2,...,N, \ell=1,2,...,p-1$, $s_{(j-1)m+i}$ and $a_{\ell,(j-1-\ell)m+i}$ denote the sign bit of the $((j-1)m+i)$-th PAM signal and the $\ell$-th amplitude bit of the $((j-1-\ell)m+i)$-th PAM signal, $\ell=1,2,...,p-1$. 

For SICM and ASCM, the symbol LLRs for SICM and LLRs of all symbols except for those corresponding to sign bits in the cas of ASCM can be expressed as follows:
\[\mathcal L(\bs v)=\sum_{i=1}^{\beta}\mathcal L(v_i,v_2,...,v_{i+\alpha-1}|\bs r),\]
where
\[ \beta=\left\{
\begin{array}{cc}
m/p &\mbox{for SICM}\\
a     & \mbox{for ASCM}
\end{array},
\right.
\]
\[ \alpha=\left\{
\begin{array}{cc}
p&\mbox{for SICM}\\
p-1&\mbox{ASCM}
\end{array}
\right.
\]
and the number of components in $\bs r$ depends on the choice of the mapping and the pair $(p,m)$.

The following two examples provide solutions for particular choices of the parameters $p$ and $m$. 

\begin{example}
Consider 4-PAM  signaling used with SICM,  and let $m=4$ as in Fig. \ref{Mappings}.
The LLR of the   $q$-ary symbol $\bs v=(v_1,v_2,v_3,v_4)$ \irina{transmitted by 4-PAM signals $y_1$ and $y_2$}   can be computed as
\begin{equation}
 \mathcal  L(\bs v=(v_1,...,v_4| r_1, r_2)=\sum_{i=1}^2 \mathcal L(v_{2i-1,}v_{2i}|r_i)\label{sicm_ex},
\end{equation} 
where 
\irina{\[ 
\mathcal L(v_{2i-1},v_{2i}|r_i)=\log  K(r_i)+\log\left(
\exp\left\{-\frac{((2v_{2i-1}-1)A (y_i) - r_i )^2}{N_0}   \right)
\right \}, i=1,2 ,
\] 
where $K(r_j)$ is a probability normalization coefficient, \irina{$A(y_i)$} denotes the amplitude of the $i$-th signal chosen according to Table \ref{Gray}.}
\end{example} 

\begin{example}
Consider 8-PAM signaling with ASCM and let $m=4$ as in Fig. \ref{Mappings}.
In this case, the LLRs for symbols associated with the sign bits are computed according to 
(\ref{LLRsbit}) and (\ref{LLRsymbol}). Let $\bs v=(v_1,v_2,v_3,v_4)$ be a code symbol 
corresponding to the amplitude bits of two 8-PAM \irina{signals}, $y_1$ and $y_2$. We assume that signs are equiprobable.  
Then the average (over equiprobable signs) conditional probabilities of amplitude bit pairs $(v_{2i-1},v_{2i})$ are equal to
\begin{eqnarray*}
P(v_{2i-1},v_{2i}|r_i)&=&K(r_i)
\left(
\exp\left\{-\frac{(\irina{A (y_i)} - r_i )^2}{N_0}   \right\} \right. \\
&+&\left. \exp\left\{-\frac{(-\irina{A (y_i)} - r_i )^2}{N_0}   \right\} 
\right)\;,  i=1,2.
\end{eqnarray*} 
The LLR for the  code symbol $\bs v$ is computed according to (\ref{sicm_ex}), where 
\[
\mathcal L\left(v_{2i-1},v_{2i}|r_i\right)=\log K(r_i) +\log \left(\sum_{j=1}^2 \exp \left\{ -\frac{((-1)^j
\irina{A(y_i)}-r_i)^2}{N_0}\right \}\right).\;
\]  
%and \irina{$A(v_{2i-1},v_{2i})$ is amplitude of the $i$-th signal chosen according to Table II. }
\end{example}
The last two examples show that for some modulation schemes, it is possible to 
compute LLRs for symbols with taking into account dependencies between reliabilities of 
bits in the binary image of this code symbol. Simulation results presented in Section \ref{Simul}  show that  
%better estimating LLRs  allows to improve the efficiency of decoding of NB LDPC codes. 
\irina{ using a mapping that better matches  the output of the NB LDPC code encoder to the PAM signal modulator/demodulator,  the decoding efficiency of the NB LDPC codes can be improved. In the next section, we compare a random coding bound on the ML decoding performance computed for the ensemble of NB LDPC codes used with QAM signaling and different mappings.}        

%\begin{figure}
%\centering
%\inputTikZ{FIGURES/mappings}
%\caption{\label{structure} Structure of the parity-check matrix after diagonalization and reordering of columns}
%\end{figure}

\section{Random coding bound for the ensemble of almost regular NB LDPC codes} \label{AlmostRegular}

The most often used lower bound on the error probability of  ML  decoding of block codes over the AWGN channel is the Shannon
bound \cite{Shannon1959}. This bound cannot be used for PAM-modulated signals since all codewords in the Shannon bound have to  be of the same energy. 
Most of upper bounds (see e.g. \cite{sason2006performance}, \cite{polyanskiy2010channel})  use the same assumption. An upper bound which is  valid for arbitrary signal sets is  presented in  \cite{herzberg1994techniques}.

\irina{Most of upper bounds are the union-type bounds and require for computation to  know the code weight enumerators. For a long LDPC code it is computationally infeasible to find the weight enumerators. An approach to avoid this problem was suggested by R. Gallager in his famous book \cite{gallager}, where he considered a random ensemble of  regular LDPC codes and derived the average over the ensemble code spectrum. Further we follow the same approach.}
 
In this section, we compute the   
Herzberg-Poltyrev (HP) upper
bound  in \cite{herzberg1994techniques} on the error probability of ML  decoding  for the  random ensemble of ``almost regular'' NB LDPC codes  in \cite{bocharova2020optimization}  used with PAM signaling.
For completeness of the paper, we present the corresponding  upper bound  in the Appendix~\ref{appA}. 

The average  binary Hamming weight spectrum for the ensemble of  almost regular NB LDPC codes   with two and three nonzero elements in each column of their parity-check matrix was derived  in \cite{bocharova2020optimization}, where a finite-length random coding  bound on ML decoding error probability of NB LDPC codes used with BPSK signaling was computed. In order to compute bound (\ref{UB}) in the case of PAM signaling it is necessary to know the SEDS of the code.
We aim at finding the relation between the Hamming weight enumerators and the SEDS for the ensemble in \cite{bocharova2020optimization}.  
 %The average  binary Hamming weight spectrum for the ensemble of  almost regular NB LDPC codes   with two and three %nonzero elements in each column of their parity-check matrix was derived  in \cite{bocharova2020optimization}, where a %finite-length random coding  bound on ML decoding error probability of NB LDPC codes used with BPSK signaling was %computed. 

%The relation of  the Hamming distance spectrum to  the Euclidean distance  spectrum for a given code ensemble depends on  indexing of signal points and mapping code symbols to
%PAM signals. The difference between code analysis for BPSK and for PAM signalling is that codes  which are linear in the Hamming space  can be not linear over the Euclidean
%space. The error probability can be different for different codewords. To overcome this problem  we  follow the approach in \cite{herzberg1994techniques}. More specifically, we  consider a chosen mapping of code symbols to PAM signals and compute the average  error probability not only over the code ensemble but also over codewords of a given code from ensemble.

%\bor{
%More specifically, for linear codes error probability is the same for all codewords. We estimate error 
%probability for all-zero codeword by upper bound which is 
%expressed through weight enumerator, which characterize pair-wise distances between all-zero word and other 
%words.  This approach does not work for coded modulation since the set of modulated codewords does not form a 
%linear space.
%

%\subsection{Computing the Euclidean distance spectra of the ensembles of almost regular NB LDPC codes }
In this section, we  survey the existing ensembles of binary and NB LDPC codes.  Then the average SEDS for the ensemble of irregular NB LDPC codes in \cite{bocharova2020optimization} used  with SICM and  BICM, ASCM, or BPCM mappings are derived via the average Hamming distance spectra of this NB code ensemble and its binary image, respectively. 

% \subsubsection{Ensembles of LDPC codes}   
\subsection{Ensembles of NB LDPC codes}

Various ensembles of irregular binary LDPC codes were studied in \cite{luby2001efficient}, \cite{burshtein2004asymptotic}, and  in \cite{litsyn2003distance}.  A generalization of the ensemble in  \cite{burshtein2004asymptotic} to an ensemble of NB LDPC   over GF$(2^m)$ determined by the ensemble of irregular bipartite graphs with given degree distributions on variable and check nodes, where each edge is labeled  by an element of GF$(2^m)$,  was studied in  \cite{kasai2011weight}.  In particular, the average  symbol Hamming weight and bit Hamming weight spectra of the random ensemble of irregular NB LDPC codes were derived. 

However, for finite-length analysis both the ensemble of irregular binary LDPC codes  in 
\cite{luby2001efficient}, \cite{burshtein2004asymptotic} and its generalization to the nonbinary case in  \cite{ kasai2011weight} have the same shortcoming. They do not determine  irregular codes with  predetermined column and row weight distributions.    
Due to unavoidable parallel edges in the code Tanner graph, 
the true degree distributions may differ from the expected one, and this phenomenon complicates the finite-length
analysis of the ensemble.
The finite-length analysis for the ensemble in \cite{litsyn2003distance} is even more difficult. Asymptotic
generating functions for code Hamming weight spectra were found in \cite{litsyn2003distance}.
Ensembles of  both binary  and  NB  regular LDPC codes  were first  analyzed by Gallager in \cite{gallager}. Later, a few different ensembles of binary LDPC codes were studied in~\cite{litsyn2002ensembles}. 

For the Gallager ensemble of binary 
$(J,K)$-regular codes,  the parity-check matrix for a code with design rate  $R=1-J/K$ 
%$R=1-r/n$ 
consists of $J$ strips
$ H_{\rm b}^{\rm T} = \left( 
H_{{\rm b},1}^{\rm T} \; |\;
H_{{\rm b},2}^{\rm T} \; \dots |\;
H_{{\rm b},J}^{\rm T}
\right) ^{\rm T}\!$, 
 where each strip $H_{{\rm b},i}$ of width $L=r/J$
is a random permutation of the first strip which can be chosen in the form 
\[
H_{{\rm b},1}=(\underbrace{    I_{L}\;...\;I_{L}\;}_{ K}),
\] 
where $I_{L}$ is the identity matrix of order $L$.

Average Hamming weight spectra for the corresponding  ensembles of regular LDPC codes were derived in  \cite{gallager} and \cite{litsyn2002ensembles}. In \cite{andryanova2009binary},  asymptotic average Hamming weight spectra for ensembles  of regular NB LDPC  codes  over GF$(2^m)$ were obtained. In \cite{bocharova2017average}, we presented a low-complexity recurrent procedure for computing exact Hamming weight spectra of both binary and NB random ensembles of regular  LDPC codes.

As was mentioned before, for NB LDPC codes over small alphabets, the average column weight of the parity-check matrix around the interval [2.2, 2.4] is preferable. This is the reason for focusing on the ensemble of ``almost regular'' NB LDPC codes in this paper.

The ensembles  of   almost regular binary and NB LDPC codes with  only two and three nonzero elements in each column of their parity-check matrices  were considered  in \cite{bocharova2020optimization}. In the same paper, the  low-complexity procedure for computing the average Hamming weight spectra in \cite{bocharova2017average} was applied to compute the corresponding spectra for  these ensembles.
The ensemble of  NB LDPC codes  in \cite{bocharova2020optimization} is based on the binary ensemble  obtained  from  the  Gallager ensemble of binary LDPC codes by allowing a given number  $K_i\le K$ 
of identity  matrices and $K-K_i$ of all-zero $L\times L$ submatrices  $\bs 0_L$  in  strips. Without loss of  generality 
the $i$th strip can be chosen 
as random permutation  $\pi_i ( H_{{\rm b},i}) $ where  $ H_{{\rm b},i}$
has the form
\begin{equation}
H_{{\rm b},i}=(\underbrace{    I_{L}\;...\;I_{L}\;}_{ K_i} \underbrace{ \bs 0_L\;...\;\bs 0_L }_{K-K_i} )\;, 
\quad i=1,...,J.
\label{checkmatr}
\end{equation}
%where $\bs 0_P$ is the all-zero matrix of order $P$.

%The other $J-1$ strips of size $M\times MK$ are random column permutations $\pi_i (\tilde  H_i) $, $i=2,...,J$,
% of \[\tilde H_1=\left(I_{M}...I_{M}\right),\]
That is, the strips in the generalized ensemble are permuted versions of Gallager's  strip with 
some identity matrices replaced by the all-zero matrices of the same order.
By  choosing $K_i$, we adjust  the column weight and row weight distributions.  
The corresponding random ensemble of NB LDPC codes is determined  by the parity-check matrix  (\ref{checkmatr}) whose nonzero entries are  labeled by randomly chosen elements of GF($2^m$). We denote the labeled parity-check matrix by $H_{\rm L}$. We refer to this NB ensemble as $\mathcal N$.

The straightforward method for constructing a  binary image of the ensemble of NB codes is replacing the elements of GF$(2^m)$ in the labeled matrix $H_{\rm L}$  by the corresponding binary $m\times m$ companion matrices of the field elements. Thus, we obtain binary strips of size $Lm\times n$. All $J$ strips of the random matrix are generated as one of $n!$ permutations  of the corresponding binary strips. We refer to this binary ensemble as $\mathcal B$.
 
The distance properties of these two binary ensembles are different, and the choice of ensemble depends on 
the mapping performed by the PAM modulator. In what follows, we analyze the ensemble $\mathcal N$ in relation to SICM mapping, and the binary ensemble $\mathcal B$ to analyzing NB LDPC codes used with BICM.  
In the next  subsections, we present  a technique for computing  the  average SEDS for the ensembles  $\mathcal N$ and $\mathcal B$ used with PAM signaling. 

\subsection{Average squared Euclidean distance spectrum  of the ensemble of  NB LDPC coded PAM signals \label{spec}}
In order to compute bound (\ref{UB}) for the aforementioned NB LDPC code ensemble used in conjunction with PAM signaling, it is necessary to know its SEDS.
There are two obstacles related to the computing of the SEDS of the coded modulation signals. First, the spectrum depends on the indexing of the signal points and the symbol-to-PAM-signal  mapping. Second, despite  the code linearity in the Hamming space, the corresponding  set of the coded modulation signal sequences does not form a linear subspace of the Euclidean space. Consequently, the modulated codewords can have different decoding error probabilities. 

The first obstacle can be overcome by deriving the SEDS via the average symbol (bit)  Hamming weight  spectrum for the ensemble $\mathcal N$ and for its binary image $\mathcal B$ in the cases of SICM and BICM, respectively.    
The second obstacle we overcome by following the approach \cite{herzberg1994techniques}. More specifically, for a chosen code symbol-to-PAM-signal mapping, averaging of the ensemble average SEDS is performed over the pairs of codewords of a given code in the ensemble.

In the calculations  below, we use notions of combinatorial and moment generating functions as well as the composition of generating functions. For completeness of the paper, the corresponding definitions and lemma are given in Appendix.  

Summarizing, we aim at deriving the average SEDS for the ensemble $\mathcal N$ and its binary image $\mathcal B$. 
They are obtained  in two steps. First, the average Hamming spectra for these ensembles are derived. Then compositions of the derived averaged Hamming weight generating functions with the moment generating function of the normalized squared Euclidean distance  (SED) between PAM signals per symbol Hamming weight or bit, are computed. 

Next, we revise the approach to calculating the precise average symbol and bit Hamming weight enumerators of the ensemble of NB LDPC codes.

\subsubsection{Average Hamming weight spectra for the ensembles $\mathcal N$ and $\mathcal B$}
The average combinatorial generating function (CGF) $ F(s)$ for the $q$-ary symbol Hamming weight enumerator of the ensemble $\mathcal N$ is derived in \cite{bocharova2020optimization}. It has the form 
\begin{equation}  F(s)=\sum_{w=0}^{N}F_{N,w}s^{w}, \label{genNB}\end{equation}
\[F_{N,w}=(q-1)^{w(1-J)}\binom{N}{w}^{1-J}\prod_{j=1}^{J}f_{j,w}^{\rm strip},\]
where, as shown in \cite{bocharova2020optimization}, the symbol Hamming weight enumerator CGF for the sequences $\bs x$ satisfying the $j$-th strip of the parity-check matrix is 
\begin{equation}
f_{j}^{\rm strip}(s)=\sum_{w=0}^{N}f_{j,w}^{\rm strip}s^{w}= 
\left(f_{j}^{\rm row}(s)\right)^{L}, \quad j=1,...,J
\label{genfstrip} 
\end{equation}
and 
\[
f_{j}^{\rm row}(s)=\sum_{w=0}^{K_j}f_{j,w}^{\rm row}=
\frac{\left(1+(q-1)s)^{K_{j}}+(q-1)(1-s)^{K_j}\right)}{q}(1+(q-1)s)^{K-K_{j}}\;\]
is the weight enumerator CGF of $q$-ary sequences $\bs x$ of length $N=n/m$ satisfying the nonzero part of one $q$-ary parity-check equation, $f_{j,w}^{\rm strip}$ is the $w$-th coefficient of the series expansion for $f_{j}^{\rm strip}(s)$.

In order to derive the average CGF $\Psi(\rho)$ for the ensemble $\mathcal B$, we compute the CGF for the number of binary sequences satisfying a system of parity-check equations of a binary strip of size $Lm\times n$ in the binary image of the parity-check matrix as composition of generating functions 
\begin{equation}
\psi_{i}^{\rm strip}(\rho)= \left.
\left(f_{i}^{\rm row}(s)\right)^{L}\right|_{s=\phi(\rho)}= \left(f_{i}^{\rm row}(\phi(\rho))\right)^{L},\quad i=1,...,J\;,
\label{genfpsi} 
\end{equation}
where $i=1,2,...,J$, and $\phi(\rho)$ as in  \cite{el2004bounds}, \cite{bocharova2017average}, denotes the moment generating function (MGF) of the nonzero $q=2^m$-ary symbol values

\begin{eqnarray}
\phi(\rho)&=&\sum_{i=1}^{m}\frac{1}{q-1}\binom{m}{i}\rho^{i}=\frac{(1+\rho)^{m}-1}{q-1} \;.
\end{eqnarray}
Similarly to (\ref{genNB}), the average CGF of the binary ensemble $\mathcal B$ is
\begin{equation}
\Psi(\rho)=\sum_{w=0}^n \Psi_w\rho^w;\quad
\Psi_w=\binom{n}{w} ^{1-J} 
\prod_{j=1}^J 
\psi_{j,w}^{\rm strip},
\label{psi}
\end{equation}
where $\psi^{\rm strip}_{j,w}$ is the $w$-th coefficient of series expansion for  $\psi_{j}^{\rm strip}(\rho)$.

It is easy to see that computing the finite-length average symbol (bit) Hamming weight spectrum for NB LDPC codes is reduced to computing coefficients of series \irina{expansion} for functions $f_{j}^{\rm strip}$  in (\ref{genfstrip}) and $\psi_{j}^{\rm strip}(\rho)$ in (\ref{genfpsi}), $j=1,2,...,J$. It can be done recursively as in \cite{bocharova2017average}. Numerical problems can be overcome by performing computations in logarithmic domain.
\subsubsection{Average squared Euclidean distance spectra of the ensembles $\mathcal N$ and $\mathcal B$}
Next, we derive the average SEDS in terms of the known functions $\tilde F(s)$ or $\Psi(\rho)$ depending on the mapping used.
  
Assume that  an $p$-bit  binary sequence  is associated with each PAM signal point in the set $\{-2^p+1,\dots,-1,1,\dots 2^p-1\}$, for example, as in Table \ref{Gray}. Similarly to the approach in  \cite{el2004bounds} and \cite{bocharova2017average}, 
we use Lemma \ref{Lemma} in order to  represent  the CGF of the average SEDS $A^{\rm symb}(\lambda)$ and $A^{\rm bit}(\lambda)$ for the ensembles $\mathcal N$ and $\mathcal B$ as a composition of the  CGF $\tilde F(s)$ and  $\Psi(\rho)$, respectively, with the MGF of the normalized SED between the PAM signal points. 
\irina{
\bor{  
%Among mappings shown in Fig. \ref{Mappings}, BICM and SICM transform a codeword to a signal sequence 
%in symbol-by-symbol fashion, while to other mappings BPCM and ASCM do this for blocks of code symbols. 
%We start analysis with single-symbol mappings and next we demonstrate how to apply this approach to 
%multi-symbol mappings.   
   }
  
%Assume that  an $p$-bit  binary sequence  is associated with each PAM signal point in the set $\{-2^p+1,\dots,-1,1,\dots 2^p-1\}$, for example, as in Table \ref{Gray}. Similarly to the approach in  \cite{el2004bounds} and \cite{bocharova2017average}, 
%we use Lemma \ref{Lemma} in order to  represent  the CGF of the average SEDS $A^{\rm symb}(\lambda)$ and $A^{\rm bit}(\lambda)$ for the ensembles $\mathcal N$ and $\mathcal B$ as a composition of the  CGF $\tilde F(s)$ and  $\Psi(\rho)$, respectively, with the MGF of the normalized SED between the PAM signal points. 
% 

\irina{We start with considering bit-to-PAM signal based mappings: BICM, BPCM, and ASCM. When BICM is used, each group of $p$ sequential bits are mapped onto one of  $2^p$-PAM signals. In the case of multi-signal mappings such as  BPCM and ASCM, we consider   groups of  $n_{\rm s}$ sequential $q=2^m$-ary symbols, that is, groups of $n_{s}m$ bits mapped onto the group of $n_{p}$  $2^p$-PAM signals,  where 
\begin{equation}n_{\rm s}m=n_{p}p=n \label{compat1},\end{equation} and $n_{\rm s}$, $n_p$ are the smallest integers satisfying (\ref{compat1}).  Denote by $\vec s=(s_1,...,s_{n_{p}})$ a vector of $n_p$ $2^p$-PAM signals. Then we compute SEDs $d^2_{\rm E}(\vec  s_i,\vec s_j)$  for all possible vector pairs $\vec s_i$, $\vec s_j$, and Hamming distance $d_{\rm H}(\vec s _{i},\vec s_{j})$ between the binary representations of $\vec s_i$ and $\vec s_j$. %Examples of the spectra for BPCM and ASCM mapping are presented in the appendix. 
}}

%\bor{ 
%Assume binary representations of non-coinciding code symbols are selected at random
%and independently with uniform distribution over the  symbols of code alphabet  and 
%indices of corresponding signal points are obtained using Gray mapping (see Table \ref{Gray}). }
%%We consider BICM and SICM used for mapping $2^m$-ary symbols onto the equiprobable  
%%$2^p$-PAM signal points, where  either $p=m$ or  $m$ is multiple of $p$. 
\irina{Consider a set of  pairs of NB LDPC coded  sequences of  $2^p$-PAM signals with one of the bit-to-PAM signal mappings corresponding to pairs of codewords $\vec v_i$, $\vec v_j$ of length $n$ bits at Hamming distance $d$.
We introduce the average over this set MGF of the SEDs between signal groups normalized per the corresponding Hamming distance   
%Over the set of  pairs of NB LDPC coded  sequences of  $2^p$-PAM signals with one of the bit-to-PAM signal mappings corresponding to pairs of codewords $\vec v_i$, $\vec v_j$ of length $n$ bits at Hamming distance $d$, we introduce the average MGF of the SEDs between signal groups normalized per the corresponding Hamming distance  
\begin{equation} \label{alpha1}
\alpha_{n,d}(\lambda)=\sum_{i=1}^{M^{n_p}} \sum_{j\ne i} %  p_{n,w}(d_{\rm H}(\bs s_i,\bs s_j)) \lambda^{\delta_{ij}},
\Pr\left \{ d^2_{\rm E}  (\bs s_i,\bs s_j),d_{\rm H} \right (\bs s_i,\bs s_j) | n,d   \} \lambda^{\delta_{ij}},
\end{equation}
%
%Under  BICM and SICM mappings, for pairs of 
%signal sequences corresponding to pairs of length $n$ codewords at Hamming distance $w$
%we derive the MGF of the SEDs normalized  per Hamming weight  in the form
%\begin{equation} \label{alpha1}
%\alpha_{w,n}(\lambda)=\sum_{i=1}^{M^{n_p}} \sum_{j\ne i} %  p_{n,w}(d_{\rm H}(\bs s_i,\bs s_j)) \lambda^{\delta_{ij}},
%\Pr( d^2_{\rm E}  (\bs s_i,\bs s_j),d_{\rm H}  (\bs s_i,\bs s_j) | n,w   ) \lambda^{\delta_{ij}},
%\end{equation}
where   
\[\delta_{ij}=\frac{d^{2}_{\rm E}(\bs s_i,\bs s_j)}{d_{\rm H}(\bs s_i,\bs s_j)},
\] 
is the normalized SED between PAM signal points at the SED  $d^{2}_{\rm E}(\bs s_i,\bs s_j)$ 
and the Hamming distance   $d_{\rm H}(\bs s_i,\bs s_j)$ 
between their binary representations. %nd  $\Pr \{t\}=\{p_{n,d}(\tau)\} $ denotes the conditional probability distribution on pairwise Hamming distances  $\tau=d_{\rm H}(\bs s_i,\bs s_j)$ given $n,d$ for length $n_sm$ binary bit sequences. }
}

\irina{For 4-PAM and 8-PAM signaling used with BICM mapping, the $\delta_{ij}$ values  are given in Table \ref{MGFA}  and \ref{MGFA8}, respectively.

For a given  Hamming distance $ d_{\rm H}  (\bs s_i,\bs s_j)$ the corresponding value of SED is a random variable 
which does not depend on $n,d$. Therefore, (\ref{alpha1}) can be expressed as 
\begin{equation} \label{alpha}
\alpha_{n,d}(\lambda)=\sum_{d_{\rm H}} 
\Pr\left( d_{\rm H}  (\bs s_i,\bs s_j) | n,d   \right)
\sum_{j\ne i} %  p_{n,w}(d_{\rm H}(\bs s_i,\bs s_j)) \lambda^{\delta_{ij}},
\Pr\left( d^2_{\rm E}  (\bs s_i,\bs s_j) | d_{\rm H}  (\bs s_i,\bs s_j)   \right) \lambda^{\delta_{ij}}.
\end{equation}
 
\begin{table} 
\centering
\caption{$d_{\rm H }(\bs s_i,\bs s_j)/(d_{\rm E}^{2}(\bs s_i,\bs s_j)$ for pairs of 4-PAM signals  
\label{MGFA}}
\begin{tabular}{|c|c|c|c|c|}
\hline
     $\vec s_i$/$\vec s_j$           & $00(-3)$ & $01(-1)$ & $11(1)$ & $10(3)$  \\ \hline
   $00(-3)$    & 0(0)   &  1(4)   &  2(16)    & 1(36)    \\ \hline 
   $01(-1)$    & 1(4)   &  0(0)   &  1(4)    & 2(16)    \\ \hline 
   $11(1) $  & 2(16)  &  1(4)   &  0(0)    & 1(4)    \\ \hline 
   $10 (3)$   & 1(36)   &  2(16)   &  1(4)    &0(0)    \\ \hline 
  \end{tabular}
\end{table}

\begin{table} 
\centering
\caption{$d_{\rm H }(\bs s_i,\bs s_j)/(d_{\rm E}^{2}(\bs s_i,\bs s_j)$ for pairs of 8-PAM signals  
\label{MGFA8}}
\begin{tabular}{|c|c|c|c|c|c|c|c|c|}
\hline
    $\vec s_i$/$\vec s_j$            
                     &$000(-7)$ & $001(-5)$ & $011(-3)$& $010(-1)$ & $110(1)$ & $111(3)$ & $101(5)$ & $100(7)$ \\ \hline
   $000(-7)$    & 0(0)        &  1(4)         &  2(16)     & 1(36)       & 2(64)       & 3(100)&2(144) & 1(196)    \\ \hline 
   $001(-5)$    & 1(4)        &  0(0)         &  1(4)       & 2(16)       & 3(36)       & 2(64) & 1(100)&  2(144)    \\ \hline 
   $011(-3) $   & 2(16)      &  1(4)         &   0(0)      & 1(4)         &2(16)        & 1(36) & 2(64 )& 3(100)    \\ \hline 
   $010 (-1)$   & 1(36)      &  2(16)       &   1(4)      &  0(0)        & 1(4)         & 2(16) &  3(36)   & 2(64) \\ \hline 
   $110 (1)$    & 2(64)      &  3(36)       &   2(16)     & 1(4)        &  0(0)        & 1(4)   &  2(16)   & 1(36)\\ \hline 
   $111(3) $    & 3(100)    &  2(64)       &  1(36)      & 2(16)       & 1(4)        &  0(0)   & 1(4)  & 2(16)\\ \hline 
   $101(5)$     & 2(144)    &  1(100)     &  2(64)      & 3(36)      &  2(16)       & 1(4)    & 0(0)   & 1(4)\\ \hline
  $100(7)$      & 1(196)    &  2(144)     &  3(100)    & 2(64)      &  1(36)       & 2(16)  &  1(4) & 0(0)\\ \hline 
  \end{tabular}
\end{table}

}

\irina{

Let us denote ${\rm Pr}\left( d_{\rm H}  (\bs s_i,\bs s_j) | n,d   \right)$  by $\{p_{n,d}(\tau)\} $, where $\tau=d_{\rm H}(\bs s_i,\bs s_j)$. 
Then, for 4-PAM signaling  with BICM and with the Gray indexing,  
from Table~\ref{MGFA} % and \ref{MGFB} 
we obtain the MGF:
\irina{
\begin{eqnarray}  \label{alpha4}    
\alpha_{n,d}(\lambda)&=& %\sum_{\gamma}S_{\gamma}^{\rm A}\lambda^{\gamma}=
p_{n,d}(1) \frac{1}{4} \left(3\lambda^4+\lambda^{36}\right) + p_{n,d}(2) \lambda^{8}. \label{bicmpam4}
% \frac{1}{6}\left(3\lambda^4+2\lambda^{8}+\lambda^{36}\right),
\end{eqnarray}}
%and for the mapping B (with Gray amplitude mapping) with $p=m/l$, $l$ is integer, 
%from Table~\ref{MGFB} we have

%\irina{
%\begin{equation} \label{4B}
%\alpha^{\rm B}(\lambda)=\sum_{\gamma}S_{\gamma}^{\rm B}\lambda^{\gamma}=\left(\frac{1}{6} \left(3\lambda^4+2\lambda^{16}+\lambda^{36}\right)\right)^l,
%\end{equation}}
%
%where $p=m/l$, $l$ is an integer, $S_{\gamma}^{\rm A}$ and $S_{\gamma}^{\rm B}$ are the average spectra of normalized SEDs for the PAM signal sets.
Similarly, for 8-PAM signaling  with the Gray mapping as in Table \ref{Gray}, we have
from Table ~\ref{MGFA8}
\begin{eqnarray} 
\alpha_{n,d}(\lambda)&=& %\sum_{\gamma}S_{\gamma}^{\rm A}\lambda^{\gamma}=
           p_{n,d}(1) \frac{1}{12} \left(7\lambda^4+3\lambda^{36} +\lambda^{100} +\lambda^{196}\right)
           \nonumber\\
           &+& p_{n,d}(2) \frac{1}{6} \left(3\lambda^8+2\lambda^{32} + \lambda^{72} \right) \nonumber\\              
           &+& p_{n,d}(3) \frac{1}{2}\left(\lambda^{12}+   \lambda^{100/3} \right).      \label{alpha8}    
\end{eqnarray}
From (\ref{alpha4}) and (\ref{alpha8}) we see that the polynomials $\alpha_{n,d}(\lambda)$ 
are sparse. For a particular mapping  it is convenient to represent them by two matrices: the matrix
of probabilities and the matrix of degrees. Collection of polynomials for BICM, BPCM, and ASCM mappings used with 4-PAM and 8-PAM signaling   is 
tabulated in Table \ref{collection} of Appendix \ref{Apspectra}.
}

\irina{
In order to compute $\alpha_{n,d}(\lambda)$,  we have to find  the probability 
distributions $p_{n,d}(\cdot)$ for the number of ones  in the binary representations of PAM signals. 
The simplest assumption is that $d$ nonzero bits are uniformly distributed over $n$ 
code bits and that 
\[
p_{n,d}(\tau)=\binom{n_sm}{\tau}\left(\frac{d}{n}\right)^{\tau} \left(\frac{n-d}{n}\right)^{n_sm-\tau}.
\]
Tighter and more complex assumption  takes into account that the sum of block weights $\tau$ is equal to $d$
and these weights are dependent, that is,    
 \[
p_{n,d}(\tau)=
\frac{\binom{n_sm}{\tau} \binom{n-n_sm}{d-\tau}  }{\binom{n}{d}}.
\]
Under condition $p \ll n$ these two approaches give the same result.  
}

\irina{The analysis of the SICM mapping is based  on the MGF of the normalized pairwise SEDs between signal points computed under assumption of the uniform distribution  of  the binary indices of the $M=2^p$-ary PAM signal points. It has the form
\[
\alpha(\lambda)=\frac{1}{M(M-1)}\sum_{i=1}^{M}\sum_{j\ne i} \lambda^{\delta_{ij}}.
\] 
For example, for 4-PAM signaling  with  SICM with the Gray indexing,  
from Table~\ref{MGFA} % and \ref{MGFB} 
we obtain the MGF:

\begin{equation} \label{4A}
\alpha(\lambda)=\frac{1}{6}\left(3\lambda^4+2\lambda^{8}+\lambda^{36}\right) .
\end{equation}
Similarly, for 8-PAM signaling  with the Gray mapping as in Table \ref{Gray}, we have
\begin{eqnarray}
\alpha(\lambda)&=&\frac{1}{28}\left(7\lambda^4+6\lambda^{8}+2\lambda^{12}+4\lambda^{32}+
3\lambda^{36}+2\lambda^{100/3}+2\lambda^{72}+\lambda^{100}+\lambda^{189}\right).  \label{8A} 
\end{eqnarray}

In order to simplify the analysis of nonlinear subspaces of coded modulation signal sequences, we introduce  {\em the  cumulative normalized  SED} between  two sequences  $\vec s_l=(s_{l1},...,s_{lN_p})$ and $\vec s_t=(s_{t1},...,s_{tN_p})$  of PAM signal points corresponding to a pair of codewords  $\vec c_l=(c_{l1},...,c_{lN})$, $\vec c_t=(c_{t1},...,c_{tN})$ at the Hamming distance $d_{\rm H}(\vec c_l,\vec c_t)=d$. It is defined as 
\irina{
 \begin{equation}  \label{delta}
 \Delta(\vec c_l,\vec c_t)=\sum_{i  \in \mathcal D_{lt}} \gamma_{i},\quad
\gamma_i=\frac{d_{\rm E}^2(s_{li},s_{ti})}{d_{\rm H}(c_{li},c_{ti})}\mbox{, } \gamma_{i}=0 \mbox{ if } c_{li}=c_{ti},
 \end{equation} }
 where  $\mathcal D_{lt}$ is a set of  
 non-coinciding bit positions  in $\vec c_l$ and $\vec c_t$. \footnote{The cumulative normalized SED is an approximation of the normalized SED between signal sequences.  It is used  in order to apply  composition of generating functions for the analysis. \irina{Although} (\ref{delta}) is not applicable for fixed codes and fixed mappings but \irina{gives  a valid approximation  on average over the product ensemble} of random codes and random mappings. }  
}
\irina{The key point here is that we ignore the  fact that the  bits in different positions make a different contribution to the SED. The reason for neglecting this fact is that we consider the average contribution of bits in different 
positions, where averaging is performed over pairs of  codewords at  a given distance $d$ under a fixed mapping. 

For codewords $\vec v_i$ and $\vec v_j$ at Hamming distance $d$ by considering $\Delta(\cdot,\cdot)$ as a sum of  $d$ i.i.d. variables $\gamma_i$ we obtain 
from (\ref{delta}) MGF for the cumulative normalized SEDs between the coded modulation sequences  corresponding to $\vec v_i$ and $\vec v_j$ as
\[
G_d(\lambda)=\alpha^d_{n,d}(\lambda) ,
\]  
where $\alpha_{n,d}(\lambda)$ is defined in (\ref{alpha}) and exemplified in  (\ref{alpha4}) and (\ref{alpha8}).

For the ensemble $\mathcal B$,  spectrum of pairwise Hamming distances is determined by GF $\Psi(\rho)$ (\ref{psi}). Therefore, 
under assumption (\ref{delta}) the corresponding SEDS is also determined by  $\Psi(\rho)$.
By composing~(\ref{psi}) and  $G_d(\lambda)$ 
we derive the  CGF of the average SEDS for BICM, ASICM, and BPCM mappings in the form 
\begin{eqnarray}
A_{n}(\lambda)&=& \sum_{ \Delta}  A_n( \Delta) \lambda^{ \Delta}=
\sum_{ \Delta } \sum_w A_n(w, \Delta) \lambda^{ \Delta}\nonumber \\ 
&=&\sum_w {\Psi}_w \sum_{ \Delta} \frac{ A_n(w, \Delta) }{{\Psi}_w } \lambda^{ \Delta} \nonumber\\ 
&=& \sum_w {\Psi}_w  \sum_{\Delta} P_w\left (\Delta\right) \lambda^\Delta= \sum_w {\Psi}_w G_w(\lambda)
\label{SEDS_bicm}, 
\end{eqnarray} 
%
%\begin{equation}
%A_{n}(\lambda)=\sum_{d}{\Psi}_d \alpha_{n,d}^{d}(\lambda)
%\label{genfB}. 
%\end{equation} 
where $A_n(w, \Delta)$ is the average number of codeword pairs at Hamming distance $w$ and  at the cumulative normalized SED  $\Delta$, 
$ P_w(\Delta)= A(w, \Delta) / {\Psi}_w$ is a probability distribution on cumulative distance $\Delta$ for a given $w$.

}
\irina{

When SICM mapping is used, we first derive the composition of the $q=2^m$-ary weight CGF $F$ for the ensemble $\mathcal N$  and the  MGF  of the $q$-ary nonzero symbol represented in the form of $m/p$ PAM signals
\begin{equation}
\theta(s)=\frac{(1+(M-1)s)^{m/p}+(M-1)(1-s)^{m/p}}{q-1}.
\end{equation}
}
\irina{By applying Lemma \ref{Lemma} in Appendix \ref{AppB}, in the case of SICM mapping we obtain the CGF for the normalized cumulative SEDS as
\begin{equation}
A(\lambda)=F(\theta(s)) = F(\theta(\alpha(\lambda))).
\label{SEDS_sicm}
\end{equation} 

}

\begin{figure}
\begin{center}
\includegraphics[width=95mm]{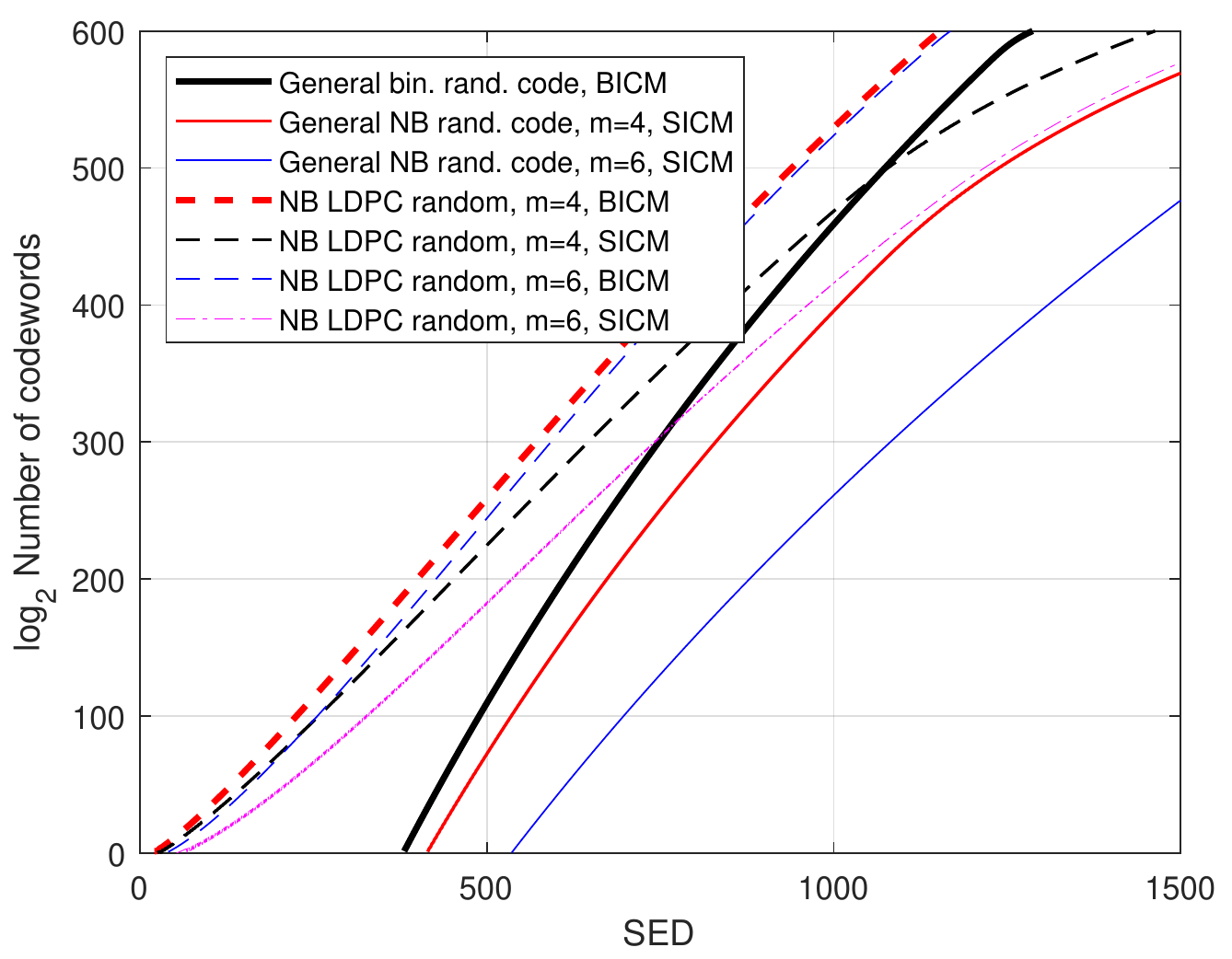}   
\caption{\label{spec225} Average cumulative normalized  SED spectrum for the ensemble of NB LDPC codes over GF$(2^4)$ and GF$(2^6)$   used with 4-PAM signaling and different mappings. The average column weight  of the base matrix is $w=2.25$}
\label{pam4m4m6w225}
\end{center}
\end{figure}

\begin{figure}
\begin{center}
\includegraphics[width=95mm]{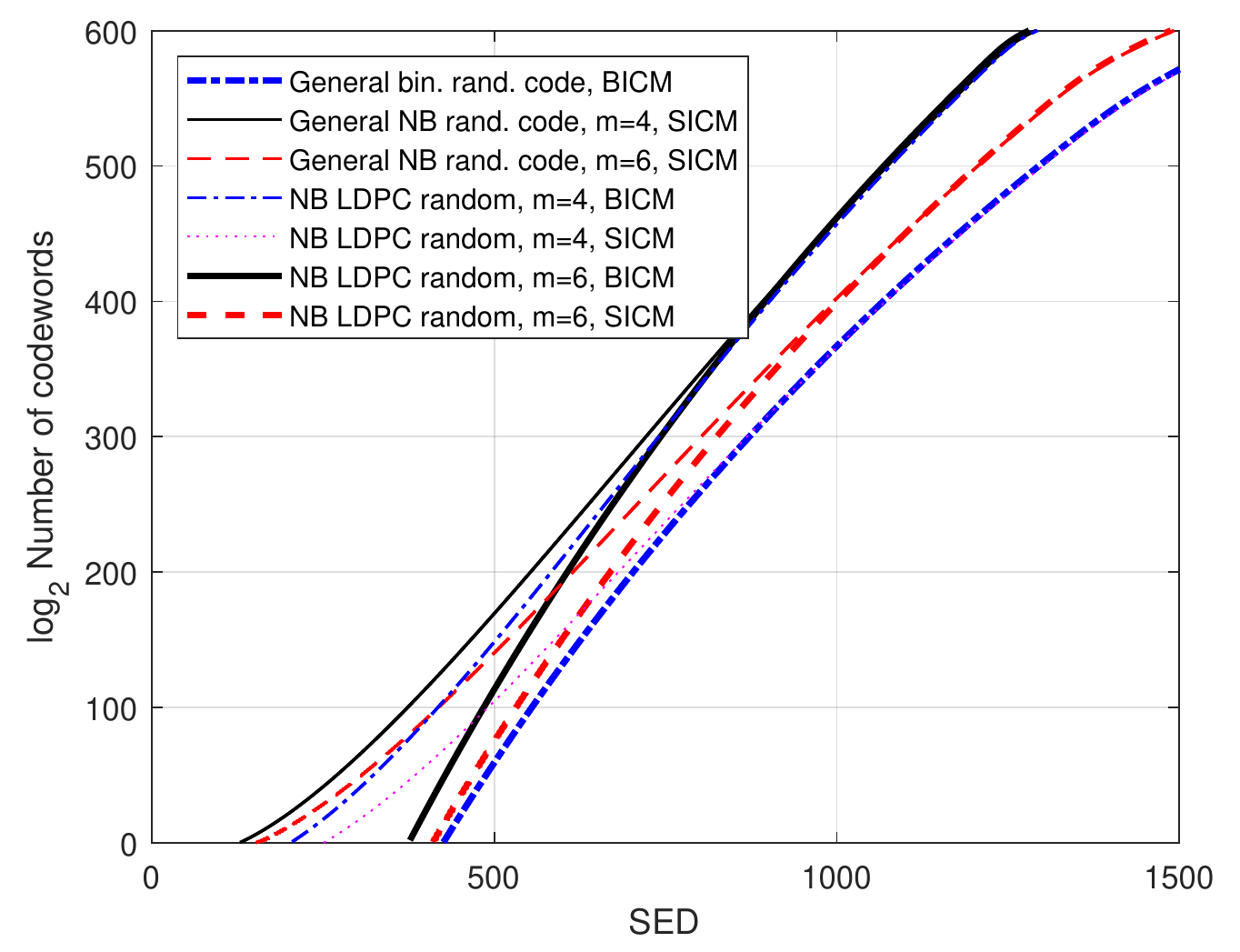}   
\caption{\label{spec283}Average cumulative normalized  SED spectrum for the ensemble of NB LDPC codes over GF$(2^4)$ and GF$(2^6)$  of the base matrix used with 4-PAM signaling and different mappings. The average column weight of the base matrix is $w=2.83$   }
\label{pam4m4m6w283}
\end{center}
\end{figure}

In Figs. \ref{pam4m4m6w225},  \ref{pam4m4m6w283} the average SEDS for the ensembles of NB LDPC codes  over GF$(2^4)$ and GF$(2^6)$ used with the 4-PAM signaling are shown for the cases $w=2.25$ and $w=2.83$, where  $w$ denotes the average column weight of the base parity-check matrix.  For comparison, the average SEDS for the ensembles of \irina{general} random binary and \irina{general} random NB codes over the same fields are presented in the same figures. \irina{Notice, that the average Hamming weight spectrum coefficients for the rate $R=(N-r)/N$ $q$-ary linear code of length $N$ are 
\[  F_{N,w}=2^{-r}\binom{N}{w}(q-1)^w.\]
If $q=2$, we obtain the average Hamming weight spectrum coefficients for the binary  random code of the same rate and length.
  }
It follows from the presented plots that SICM provides better SEDS than BICM independently of the value of $m$. As expected, the SEDS of the NB LDPC codes from the ensemble are approaching the SEDS of general random NB codes when  $w$   grows.

\begin{figure}
\begin{center}
\includegraphics[width=95mm]{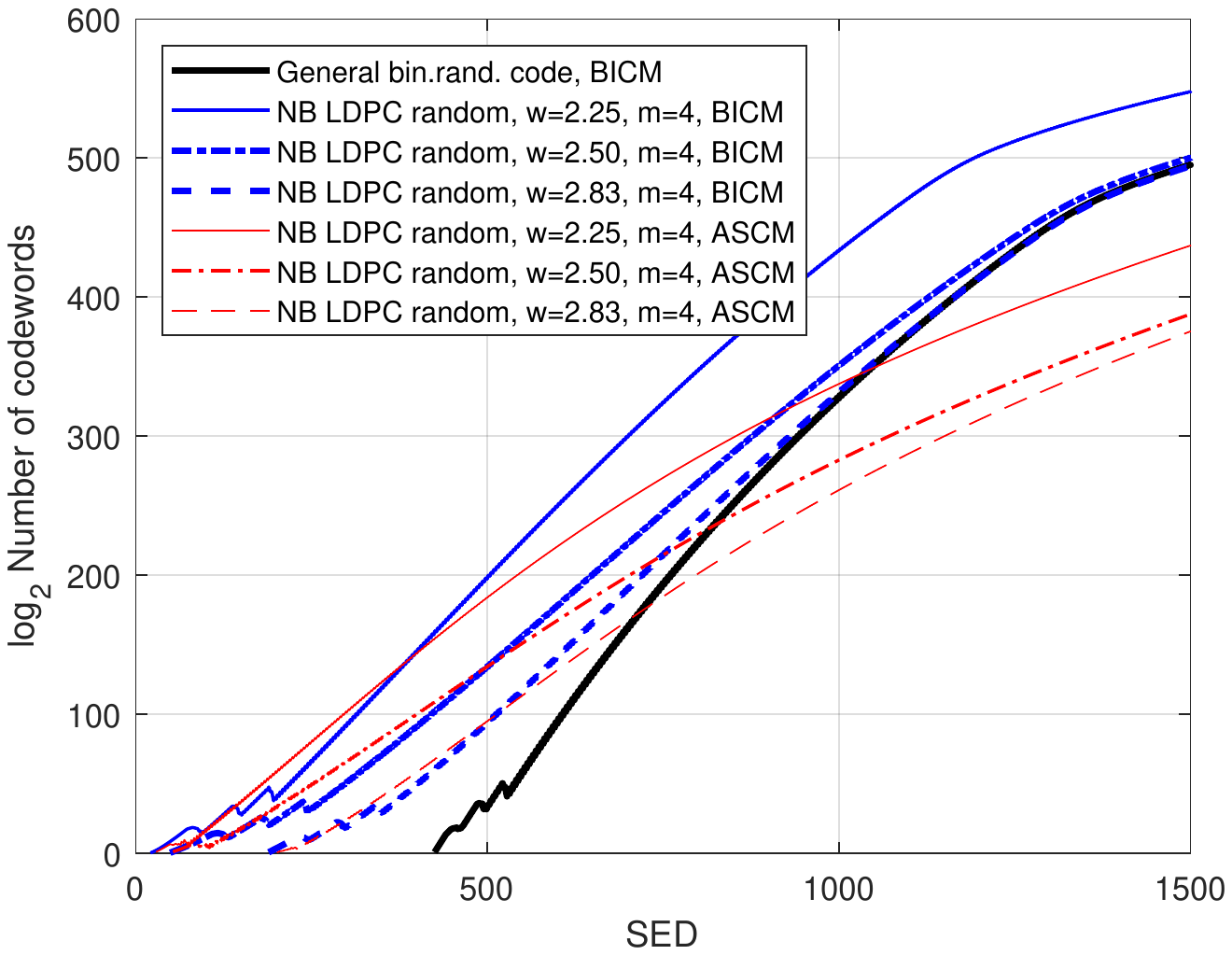}   
\caption{\label{spec283}Average cumulative normalized  SED spectrum for the ensemble of NB LDPC codes over GF$(2^4)$  of the base matrix used with 8-PAM signaling and BICM and ASCM mappings.  }
\label{pam8m4ascm}
\end{center}
\end{figure}

\irina{
In Fig. \ref{pam8m4ascm}, we present the average SEDS for the ensemble of NB LDPC codes with $w \in \{2.25; 2.5; 2.83\}$ over GF$(2^4)$ used with 8-PAM signaling. The average SEDS for ASCM mapping  are compared to the average SEDS for BICM mapping used with the same code ensembles.  In the same figure, the average SEDS for the general binary random code used with 8-PAM signaling and BICM mapping is shown. It follows from the  presented results, that ASCM SEDS are better   then the corresponding BICM spectra.  
}

In the next section, we present the random coding bound on the ML decoding error probability obtained by substituting spectra (\ref{SEDS_bicm}) and (\ref{SEDS_sicm}) to the HP upper bound  (\ref{UB}).

\section{Numerical results\label{Simul}} 

In this section, first, we compute the random coding bounds on the ML decoding error probability for the ensemble of NB LDPC codes over GF$(2^4)$ and GF$(2^6)$  used  with 4-PAM and 8-PAM signaling and different mappings. These bounds are obtained by substituting the average spectra (\ref{SEDS_bicm}) and (\ref{SEDS_sicm}) into the HP upper bound  (\ref{UB}). Comparison of these bounds with  the bound (\ref{UB}) calculated  for the \irina{general} binary  random linear code  and the \irina{general} NB random linear code  of the same length is performed. 
Then we present the simulation results for the FER performance of BP decoding  for the NB QC-LDPC codes  over GF$(2^4)$ and GF$(2^6)$,  optimized as in \cite{bocharova2020optimization} and \cite{Boch2016}. We compare the FER performance of BP decoding with the random coding bound on ML decoding error probability for random almost regular NB LDPC  codes of the same length and alphabet  sizes.

In the sequel, we use notation SNR for signal-to-noise ratio per signal component measured in dB, $w$ denotes the average column weight, $J$ and $K$ denote the maximal number of nonzero elements in each column and each row of the parity-check matrix, respectively.  
\subsection{Calculation of  random coding  bounds} 

We analyze the  ensemble of random almost regular rate $R=3/4$ NB LDPC codes  determined by the  base matrix
of size  $3 \times 12$. In  all examples, parity-check matrices have row weights 
$K_1\le K, K_2=K_3=K$.  The average column weight is computed as  $w=(24+K_1)/12$.  

\begin{figure}
\begin{center}
\includegraphics[width=90mm]{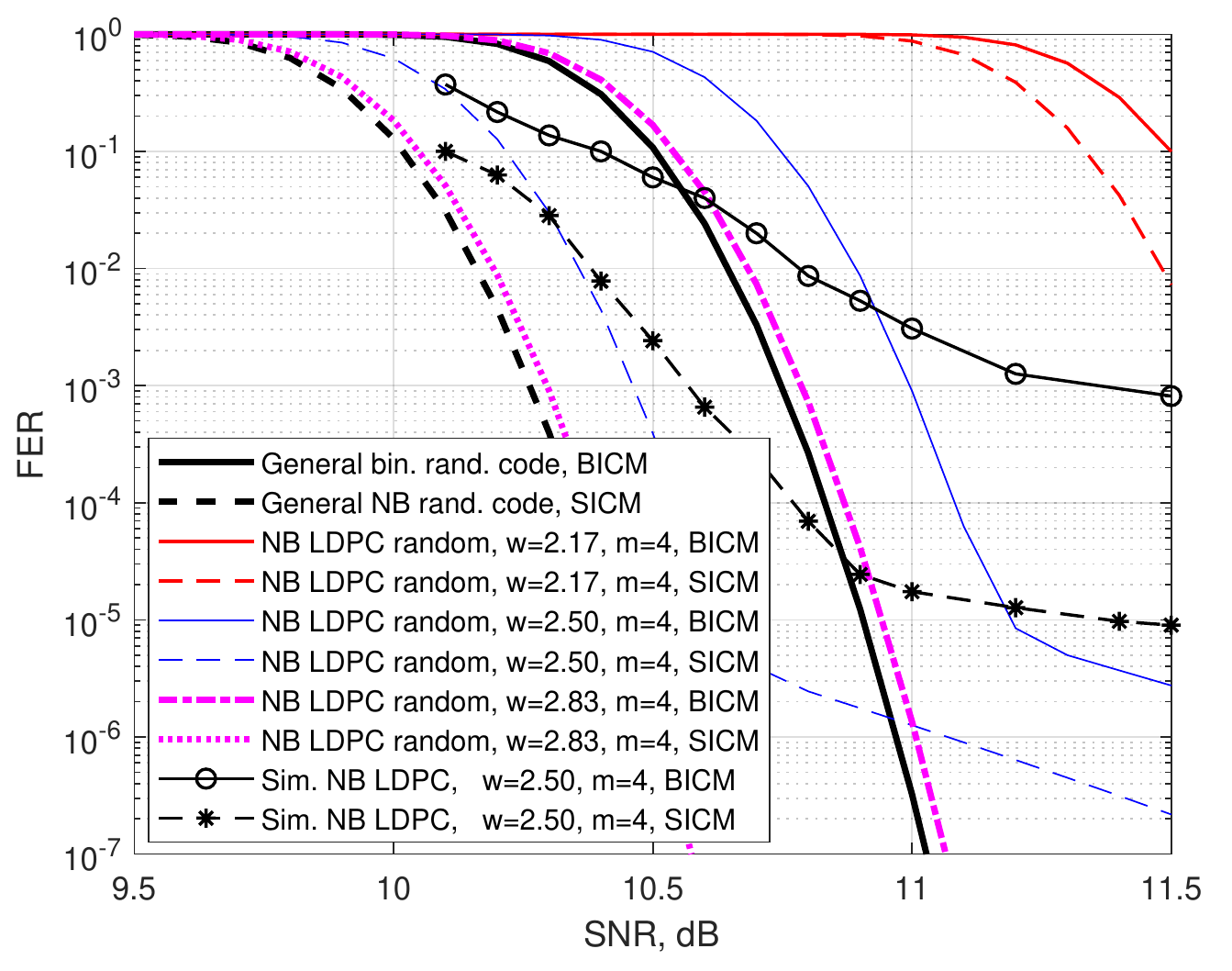}   
\caption{\label{bpsk44} Bounds on the error probability of  the ML decoding  for rate $R=3/4$ NB  LDPC codes over GF$(2^4)$  of length about 2000 bits with 4-PAM signaling. \irina{BICM limit is equal to 9.304 dB. }
}
\label{pam4m4b}
\end{center}
\end{figure}

\begin{figure}
\begin{center}
\includegraphics[width=90mm]{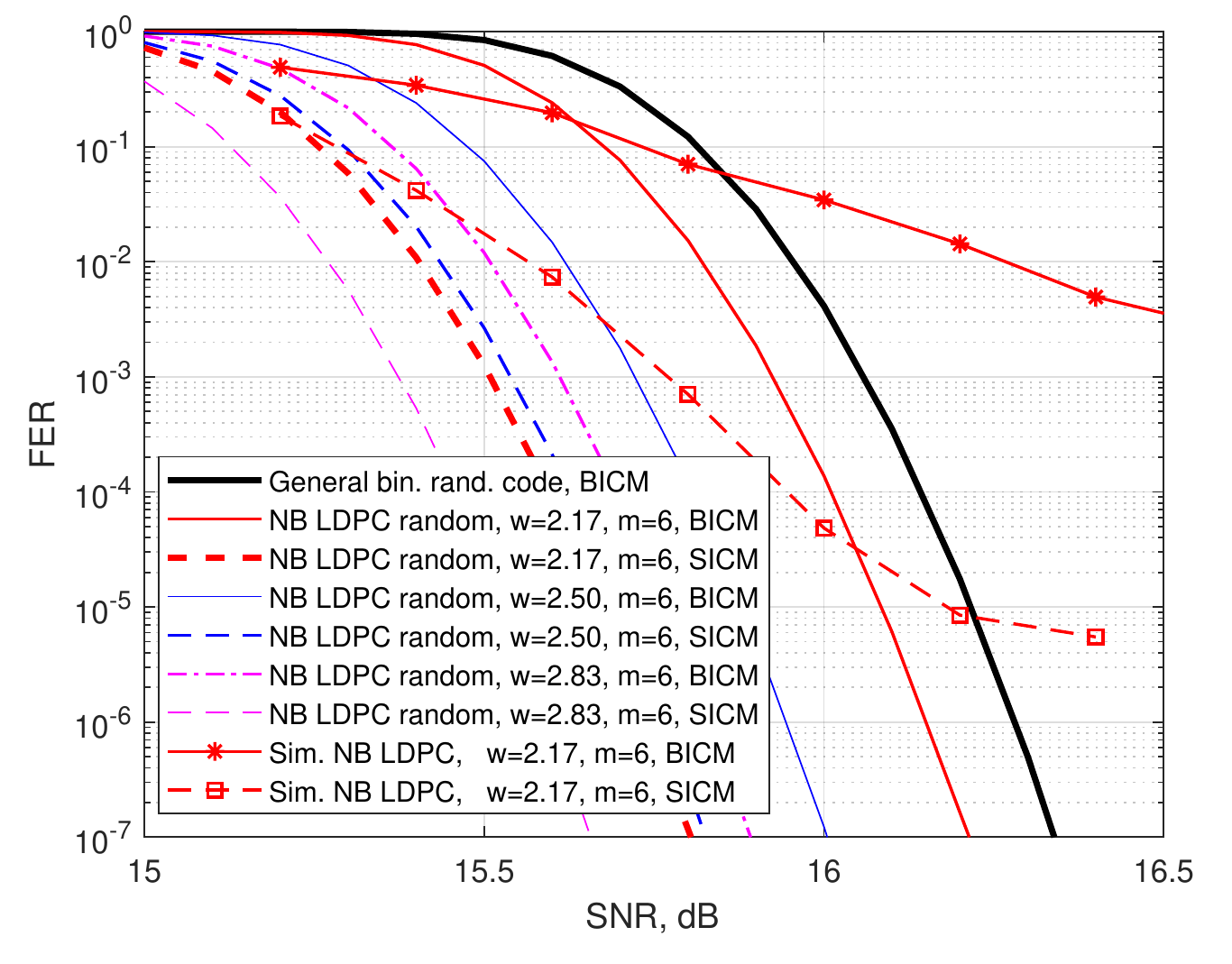}   
\caption{\label{bpsk68} Bounds on the error probability of  the ML decoding  for rate $R=3/4$ NB  LDPC codes over GF$(2^6)$  of length about 2000 bits  with 8-PAM signaling. \irina{BICM limit is equal to 14.365 dB.} }
\label{pam8m6b}
\end{center}
\end{figure}

The  bounds on the error probability of ML decoding for \irina {NB  random} LDPC codes as well as \irina{ the general  binary  and NB  random  linear codes}   are plotted in Figs. \ref{pam4m4b} -- \ref{pam8m6b}
for different alphabet sizes and modulation orders. For comparison in the same figures, we plotted the FER performance of NB QC-LDPC codes of the same length and alphabet size.

\begin{figure}
\begin{center}
\includegraphics[width=95mm]{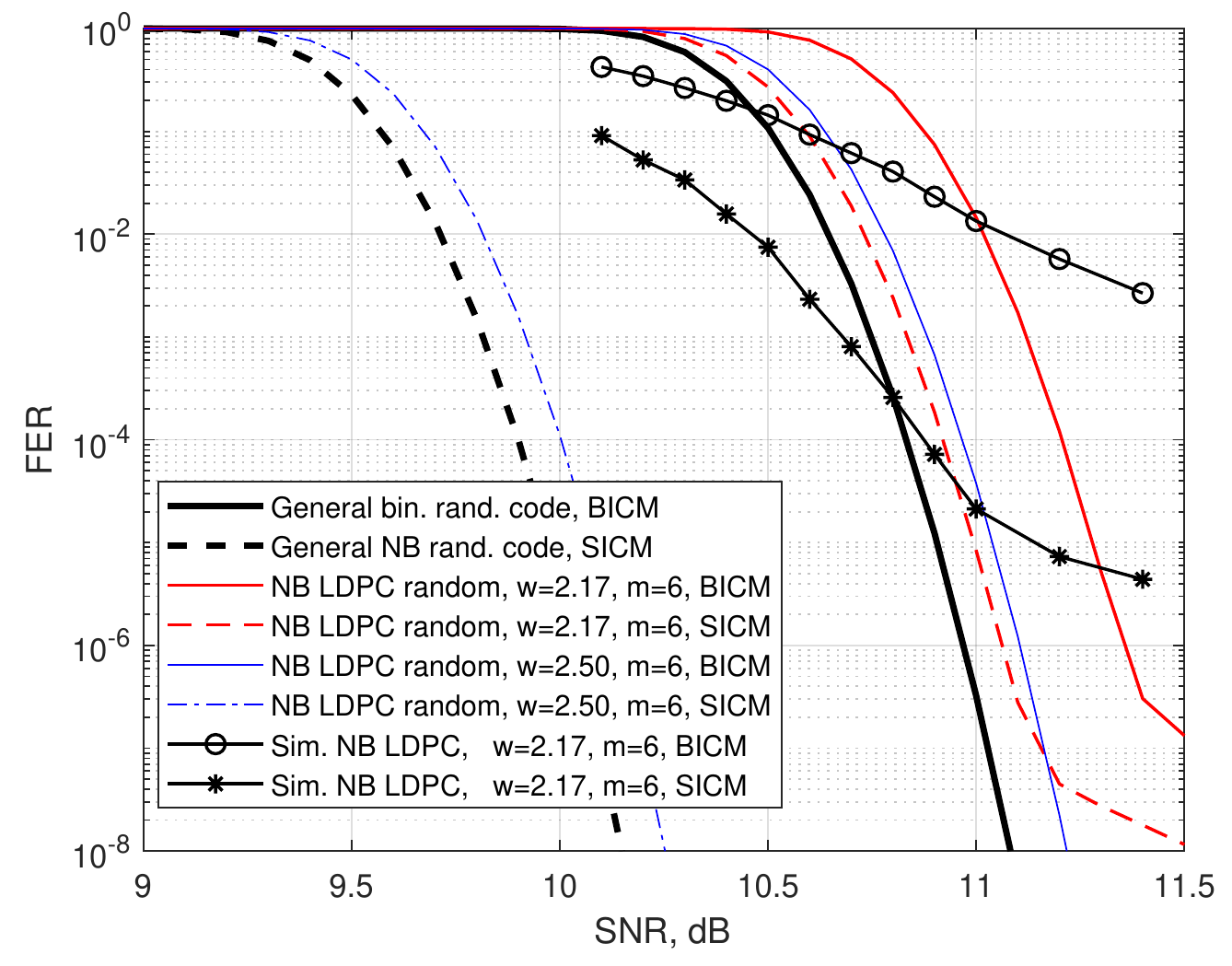}   
\caption{\label{bpsk64} Bounds on the error probability of  the ML decoding  for rate $R=3/4$ NB  LDPC codes over GF$(2^6)$  of length about 2000 bits  with 4-PAM signaling. BICM limit is equal to 9.304 dB. }
\label{pam4m6b}
\end{center}
\end{figure}

\begin{figure}
\begin{center}
\includegraphics[width=90mm]{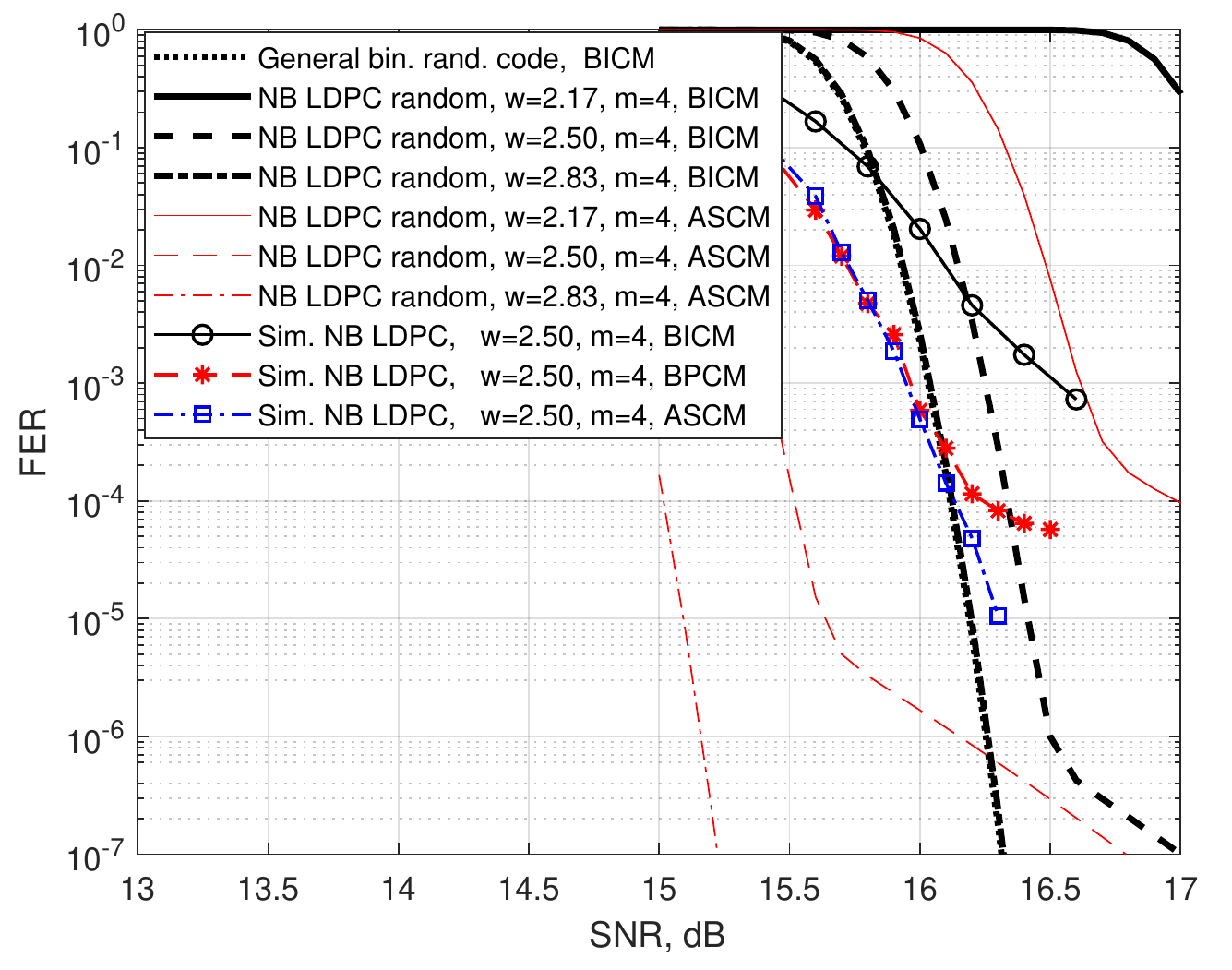}   
\caption{\label{bpsk64} Bounds on the error probability of  the ML decoding  for rate $R=3/4$ NB  LDPC codes over GF$(2^4)$  of length about 2000 bits  with 8-PAM signaling. \irina{BICM limit is equal to 14.365 dB.} }
\label{pam4m6b}
\end{center}
\end{figure}

The  presented bounds show that  
\begin{itemize}
\irina{
\item 
ML decoding error probability bounds for different parameters of the NB LDPC code ensemble and different  mappings
demonstrate the advantage of SICM and ASCM mappings over the BICM mapping. The presented bounds  show error floor phenomena for some scenarios and  demonstrate performance behavior depending on the average column weight $w$. 
The predicted behavior matches  the simulated results for the NB QC-LDPC codes with the same alphabet size and the same average column weight $w$, although, numerically the simulation results for the BP decoding  performance differ 
from the predicted ML decoding performance.  Thus, the approach based on the  
normalized SEDS can be considered  as a useful technique for the analysis of practical coded modulation systems.
}

\item {Unlike the FER performance of  \irina {general binary and NB random  linear codes}, the FER performance of NB LDPC codes over GF$(2^4)$ and GF$(2^{6})$ used with both 4-PAM and 8-PAM signaling have severe error  floors independently on the used mapping. \irina{Increase in the average column weight $w$  lowers the error floor level}}.
\item{ Both the bounds and the simulation results suggest that in the waterfall region, the FER performance of the NB LDPC codes used with  BICM is significantly worse than the FER performance of NB LDPC codes with SICM independently of the alphabet size and modulation order.  Reduction in the average column weight $w$ monotonically  worsens the FER performance of the ML decoding.}  
\irina{
\item{For the sets of coded modulation system parameters  where SICM mapping cannot be applied, ASCM and BPCM mappings provide  the ML decoding performance which is superior to the case when BICM mapping  is used.}} 
\irina{\item
Random coding bounds for ASCM and BPCM  used  in the 8-PAM scenario with $m=4$ numerically coincide with each other
(see Appendix \ref{Apspectra}). However, when used with BP decoding, ASICM provides slightly  better results than BPCM.   } 

%\item{Random NB LDPC codes over GF$(2^4)$ used in conjunction with 4-PAM and BICM have the FER performance of the ML decoding, which is close to that of the general binary \irina{random} linear codes, while the FER performance of the NB \irina{random} LDPC codes over GF$(2^4)$ used with SICM has a gap of the order of 0.2 -- 0.4 dB with respect to the FER performance of the  general NB random linear codes. } 
\irina{
\item{When increasing $m$ and $w$,  NB random LDPC codes used with  BICM  and SICM provide the ML decoding FER performance rather close to that of the  binary random linear  and NB random linear codes, respectively.}}
\end{itemize}

\subsection{NB QC-LDPC codes with  QAM signaling}
It is well known that due to their structural properties, QC-LDPC codes are  widely used  in modern communication standards such as WiFi, WiMAX, DVB-S2, and 5G standards. In this subsection, we present simulation results of the BP decoding for moderate length (about 2000 bits) optimized NB QC-LDPC codes  used with PAM signaling and different mappings. Comparison with the derived bounds as well as with the BP decoding performance for best known binary and NB LDPC codes with the same parameters  are performed. 

It is expected that sparser codes are weaker in the sense of ML decoding performance. However, 
sparseness is important for improving the performance of BP decoding. The goal of our computations 
and simulations is to evaluate a sparsity factor which allows to stay close to optimal codes
in the sense of ML decoding and to improve as much as possible BP decoding performance.     
      
In the following,  we consider a set of  rate $R=3/4$  NB QC-LDPC codes with maximum column weight of their parity-check matrices $J=3$ optimized by using techniques in \cite{bocharova2020optimization} and \cite{Boch2016}.  NB QC-LDPC codes of length about 2000 bits  with different average column weight $w$ of their base parity-check matrices  are studied. All considered NB QC-LDPC codes are determined by the base matrix of size $14\times 56$ with column weights two and three, the lifting  factor $L$ is chosen to be equal to 9, 8, and 6 for the code over GF($2^4$), GF($2^5$), and GF($2^6$), respectively. The average column weight $w$  takes on values from the set $\{2.17, 2.27, 2.35, 2.50, 2.67, 2.71\}$. These codes  used with 4-PAM and 8-PAM signaling were simulated over the AWGN channel.  The FER performance of sum-product BP decoding was simulated with a maximum of 100 iterations until 50 block errors. 

\begin{figure}
\begin{center}
\includegraphics[width=90mm]{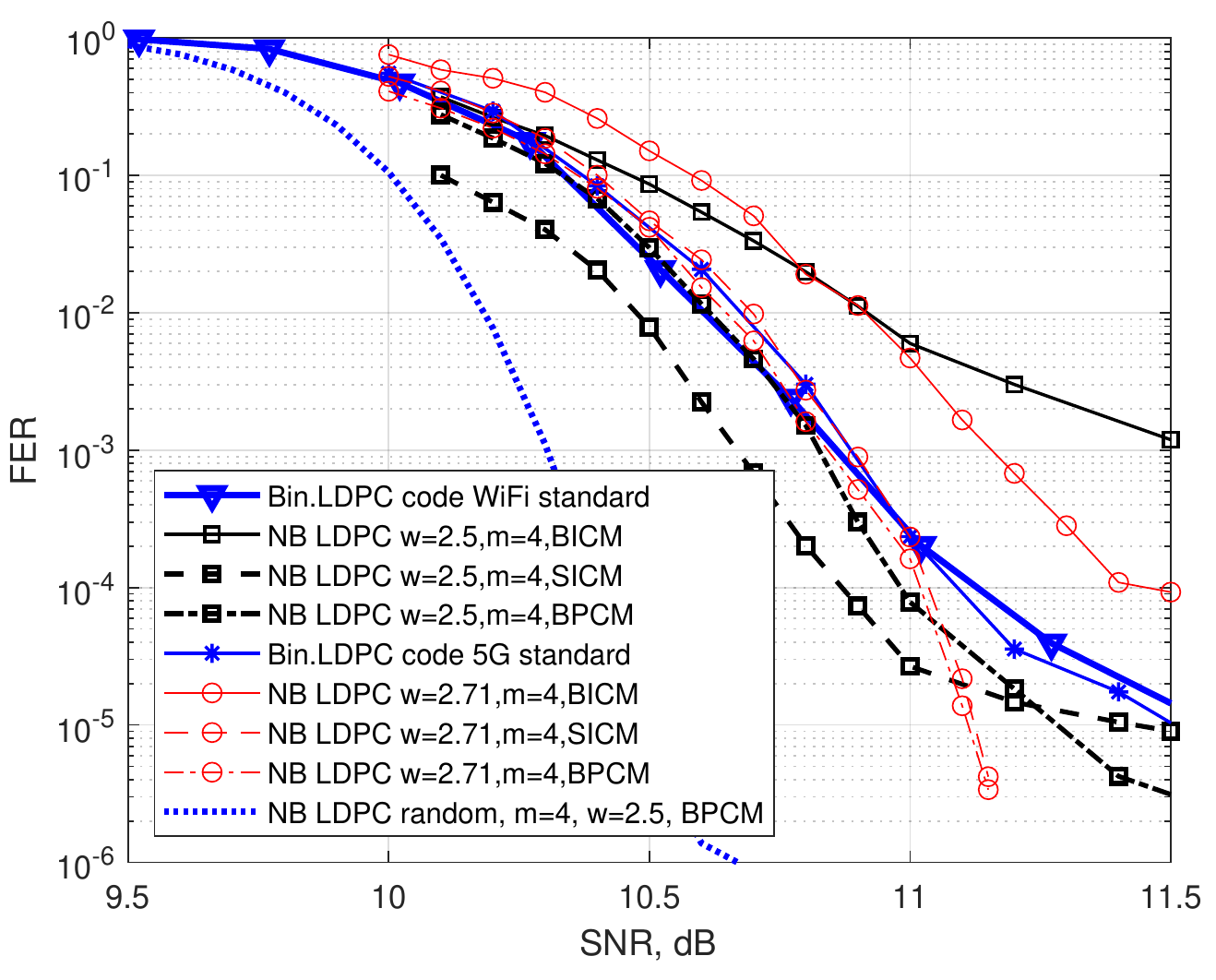}   
\caption{\label{sm4} FER performance  of the BP decoding for rate $R=3/4$ NB QC LDPC codes of length 2016 bits over GF$(2^4)$ with 4-PAM signaling. 
\irina{BICM limit is equal to 9.304 dB.} }
\end{center}
\end{figure}

\begin{figure}
\begin{center}
\includegraphics[width=90mm]{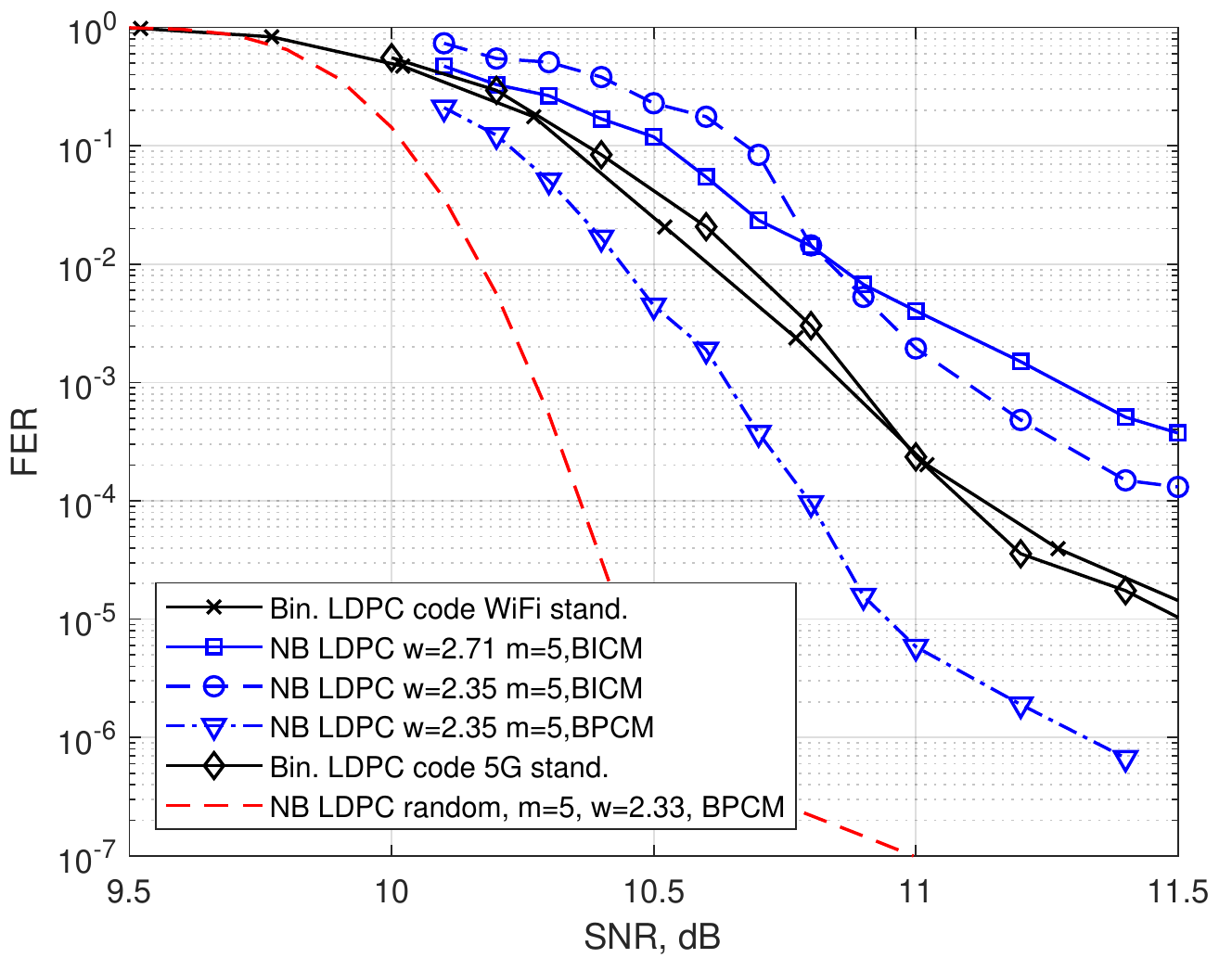}   
\caption{\label{sm5} FER performance  of the BP decoding for rate $R=3/4$ NB QC LDPC codes of length 2240 bits over GF$(2^5)$ with 4-PAM signaling.
\irina{BICM limit is equal to 9.304 dB. }}
\end{center}
\end{figure}

\begin{figure}
\begin{center}
\includegraphics[width=90mm]{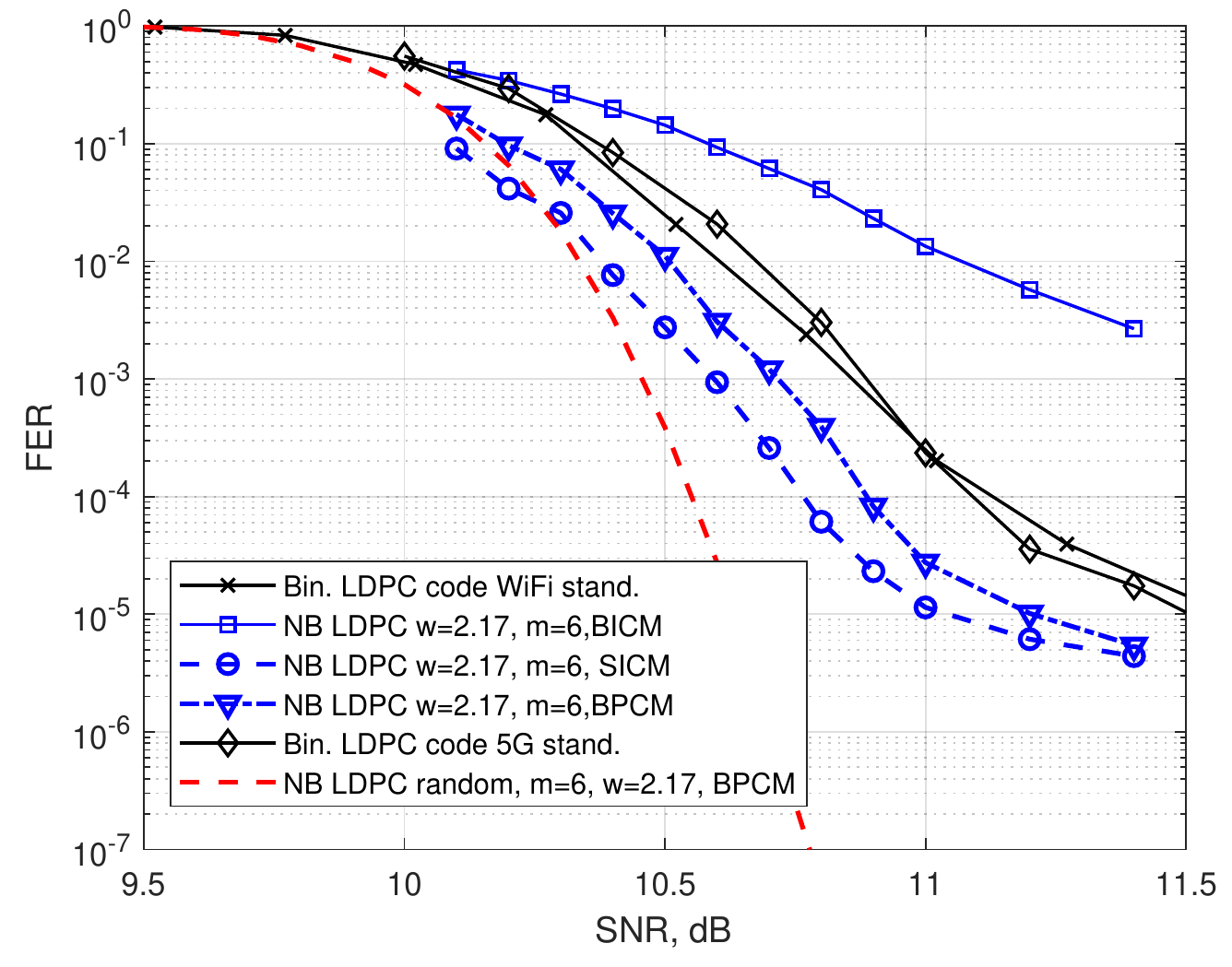}   
\caption{\label{sm6} FER performance  of  the BP decoding for rate $R=3/4$ NB QC LDPC codes of length 2016  bits over GF$(2^6)$ with 4-PAM signaling.
\irina{BICM limit is equal to 9.304 dB.} }
\end{center}
\end{figure}

The obtained FER performance of the BP decoding when using different demodulators for the field extension order $m=4$, $m=5$, and $m=6$, respectively,  and for the 4-PAM signaling  is shown in Figs. \ref{sm4} -- \ref{sm6}.  Comparison with the standard binary code in the WiFi standard as well as with the binary code in 5G standard is made.  \irina{In the same figures we showed the ML decoding error probability bound for the NB random LDPC code with parameters $m$ and $w$  which provide the best FER performance of the sum-product decoding. All figures are accompanied by a value of the BICM limit. }

It is easy to notice that the FER performance of the BP decoding for NB  QC-LDPC codes  in a large degree depends on the chosen mapping.   The NB QC-LDPC codes used with BICM  show the FER performance, which is inferior to  the corresponding performance of  the binary codes. Both SICM and  BPCM mappings provide a gain of about 0.5 dB with respect to the binary case. Increasing the number of weight three columns in the base matrix of the considered NB QC-LDPC codes, as expected,  improves the FER performance of BP decoding in the error floor region but worsens the FER performance at low SNRs independently on the mapping used.  In Fig. \ref{sm5}, the FER performances of two NB QC-LDPC codes ($w=2.35$ and $w=2.71$) used with 4-PAM and BICM  are shown. The NB QC-LDPC code with $w=2.71$ outperforms  the NB QC-LDPC code with $w=2.35$ in the error floor region but loses at low SNRs.  \irina{It follows from the presented plots that the ML decoding error probability bound mimics the behavior of the BP decoding FER performance. The observed gap between the bound and the simulation results is about   0.2--0.5 dB at the FER $\approx 10^{-4}$.}

\begin{figure}
\begin{center}
\includegraphics[width=85mm]{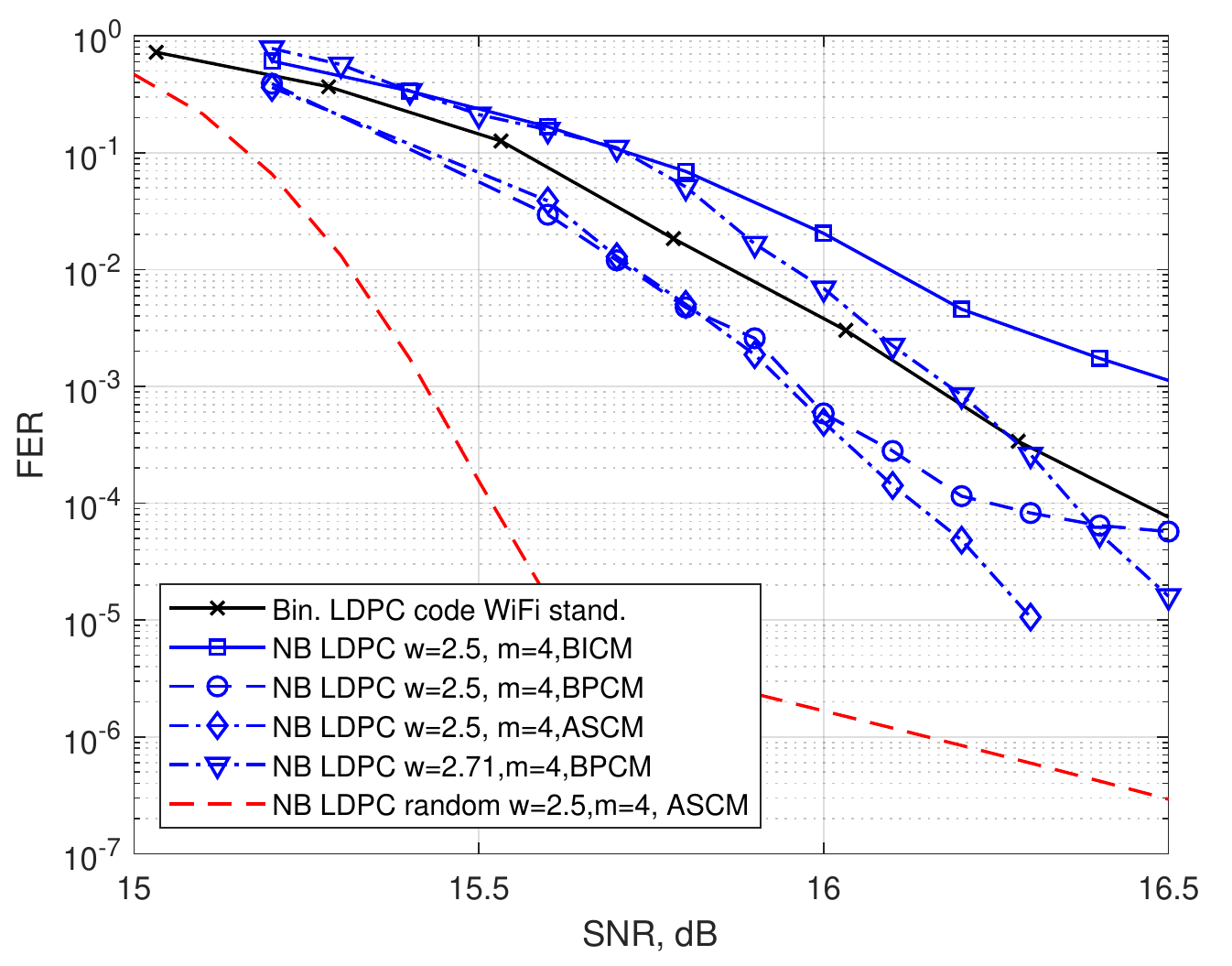}   
\caption{\label{sm4q64} FER performance  of the BP decoding for rate $R=3/4$ NB QC LDPC codes of length 2016  bits over GF$(2^4)$ with 8-PAM signaling.
\irina{BICM limit is equal to 14.365 dB.} }
\end{center}
\end{figure}

\begin{figure}
\begin{center}
\includegraphics[width=85mm]{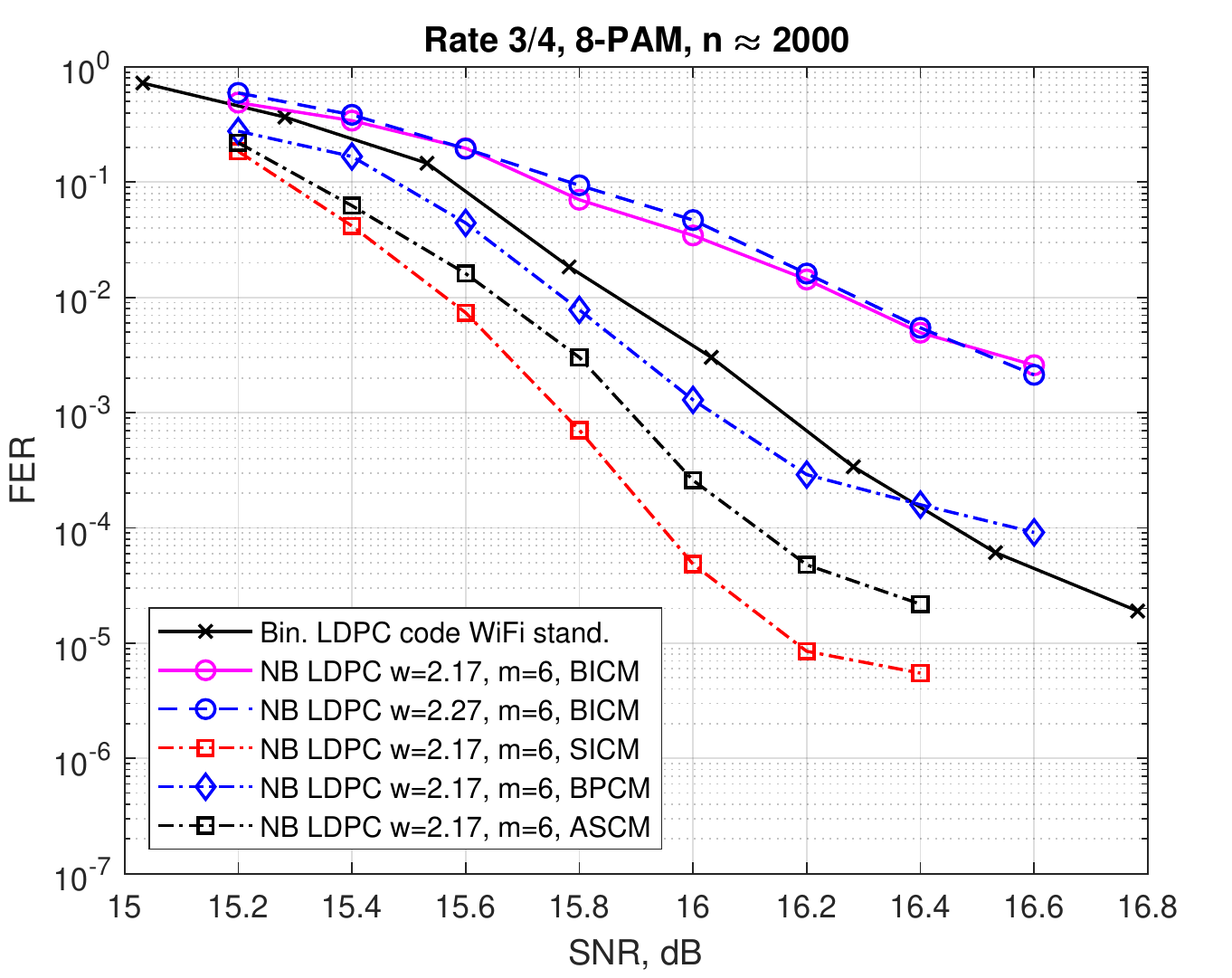}   
\caption{\label{sm6q64} FER performance  of the BP decoding for rate $R=3/4$ NB QC LDPC codes of length 2016  bits over GF$(2^6)$ with 8-PAM signaling.
\irina{BICM limit is equal to 14.365 dB.} }
\end{center}
\end{figure}

Simulation results for NB QC-LDPC codes with 8-PAM signaling and field extensions $m=4$ and  $m=6$ are presented in Figs. \ref{sm4q64}--\ref{sm6q64}. Similarly to the 4-PAM case, NB QC-LDPC codes with BICM \irina{exhibit FER performance which is inferior to  that of the binary code. }
NB QC-LDPC codes with SICM and   ASCM  demonstrate a 0.5 dB gain over the binary code. Comparison of the performance curves for the NB codes over GF$(2^6)$ with $w=2.17$ and $w=2.27$  in Fig. \ref{sm6q64} shows that increasing the number of weight three columns in the code base matrix  if 8-PAM signaling is used with BICM  does not improve performance  significantly.   


Notice  that the obtained results are consistent with conclusions drawn in \cite{declercq2004regular} and \cite{rezqui} for regular NB LDPC codes. In particular, BICM used with both 4-PAM  and 8-PAM signaling  loses about 1 dB compared to SICM used with the same signaling.  
\irina{Comparison with the bound shows a gap of 0.1--0.6 dB at the FER $\approx 10^{-4}$.}
%
%In Fig. \ref{comp}, we  compare the BER performance of our optimized NB QC-LDPC codes used with BICM and SICM
%with the same performance of the regular NB LDPC codes in \cite{declercq2004regular}. We conclude that the optimized NB QC-LDPC codes outperform the longer regular NB LDPC codes in \cite{declercq2004regular}, SICM mapping provides better performance than BICM mapping. 
%\begin{figure}
%\begin{center}
%\includegraphics[width=85mm]{comparison.pdf}   
%\caption{\label{comp} Comparison of BER performance  of BP decoding for rate $R=3/4$ NB QC LDPC codes of length 2016  bits over GF$(2^6)$ with 4-PAM signaling  and  the same performance of length 2880 NB LDPC codes over GF$(2^6)$ in \cite{declercq2004regular}}
%\end{center}
%\end{figure}
\section{Conclusion}
NB QC-LDPC codes used with QAM signaling over the AWGN channel were analyzed.  Four different mappings of code symbols to QAM signals were studied.  It was shown  that  BICM used with both 4-PAM and 8-PAM signaling is inferior to SICM with the same signaling by about 1 dB. 

The NB QC-LDPC codes used in conjunction with BICM exhibit \irina{the FER performance} which is inferior to that of the binary counterparts.
More sophisticated mappings could provide for better performance compared to the BICM. If the modulation order is not a divisor of the alphabet order, then the BPCM and ASCM outperform the BICM by about 0.5 dB.  Yet, these mappings do not improve the performance with respect to SICM when the modulation order is not a divisor of the alphabet order. 

The random ensemble of almost regular NB LDPC codes used with QAM signaling was analyzed.  The average squared Euclidean distance spectra for codes in this ensemble used with \irina{different} mappings were derived.
By substituting the computed average squared Euclidean distance spectra  to the Herzberg-Poltyrev upper bound, the  tightened  finite-length upper bounds on the error probability of ML decoding  for irregular NB LDPC codes over GF($2^m$) were derived.
\irina{The obtained random coding bound mimics the behavior of the FER performance of BP decoding of practical NB QC-LDPC codes. In particular, in some scenarios it exhibits a severe error floor independently of the mapping used. The gap between the bound and the simulation results varies from 0.1 to 0.6 dB depending on the alphabet size, code density, and modulation order.}     
 % The obtained random coding bound  exhibits a severe error floor independently of the mapping used. %However, the optimized NB QC-LDPC codes with BP decoding have much lower error floors than those predicted by the %random coding bounds since their distance properties were optimized by adjusting  the average column weight of their %parity-check matrices and choosing  proper labeling.
The presented simulation results and comparisons with the bounds on error probability showed that NB QC-LDPC codes outperform  known binary QC-LDPC codes. 
    
\begin{appendix}

\subsection{Upper bound \label{appA}}
The Shannon-Herzberg-Poltyrev  \cite{herzberg1994techniques} is  valid for arbitrary signal sets of length $n$ signals
 \begin{eqnarray}
P_e &\le&  
\sum_{w\le w_0} S_w \Theta_w(x)
+  1-\chi_{n-1}^2\left(\frac{w_0}{\sigma^2}\right).
\label{UB}
\end{eqnarray}
Here %$f(x)=(1/\sqrt{2\pi})\exp{-x^2/2}$ is the  Gaussian probability  density function,  
%$Q(x)  =  \int_{x}^{\infty} f(x) dx$, 
\begin{eqnarray*}
\Theta_w(x) & = & 
 \int_{\sqrt{w}/2}^{\sqrt{w_0}}
f\left(\frac{y}{\sigma}\right)\phi_{n-1}
\left(\frac{w_0-y^2}{\sigma^2}\right)
dy \; 
\end{eqnarray*}
and
\begin{eqnarray*}
f(x)&=&\frac{1}{\sqrt{2\pi}} \exp\left\{-\frac{x^2}{2}\right\}\;,  \\
\phi_n(x)&=&\frac{x^{n/2-1}e^{-x/2}}{2^{n/2}\Gamma(n/2)}\;, \\% \mbox{ and}\\
\chi_n^2(x)&=&\frac{\gamma(n/2,x/2)}{\Gamma(n/2)}\;
\end{eqnarray*}
%w_0&=&\left \lfloor \frac{r_0^2n}{r_0^2+n}\right \rfloor ,
%r_x=r_0\left(1-\frac{x}{\sqrt n}\right) \; ,\\
%\mu_w(r)&=&\frac{1}{r}\sqrt {\frac{w}{1-w/n}}, 
%\beta_w(x)=\left(1-\frac{x}{\sqrt n}\right)\sqrt {\frac{w}{1-w/n}} \; , 
%\end{eqnarray*}
are Gaussian probability  density function, probability density function and probability distribution function of the $\chi$-squared distribution with $n$ degrees of freedom, respectively,
%v$f(x)=(1/\sqrt{2\pi})\exp{-x^2/2}$ denotes the  Gaussian probability  density function, 
$S_{w}$ is the $w$-th spectrum coefficient,  and 
parameter $w_0$ is a solution of the equation
\begin{equation}\label{UB_eq}
\sum_{w\le w_0}S_w \int_0^{\arccos \sqrt {\frac{w}{4w_0}}} \sin^{n-3}\phi \quad d \phi =\sqrt{\pi}
\frac{\Gamma\left(\frac{n-2}{2}\right)}{\Gamma\left(\frac{n-1}{2}\right)} \; .
\end{equation}
\subsection{Generating functions. \irina{Compositions of generating functions} \label{AppB}}
The {\em combinatorial generating function} (CGF) $\psi(s)$ for a sequence of \irina{nonnegative} numbers $u_0,u_1,\dots$ 
is defined as 
\irina{\[\psi(s) =\sum_{i=0}^\infty u_is^i,\] }
where $s$ is a formal variable. 
\irina{We interpret $u_i$ as a multiplicity of occurrences of value $i$ in the same realization.} 
If the sequence $\bs u$ is random then  the {\em average combinatorial generating function} $\tilde \psi(s)$ of $\bs u$ is obtained by averaging $\psi(s)$ over all possible vectors $\bs u$. 
   
The {\em  moment generating function} (MGF) $\theta(\rho)$  for a discrete 
random variable $v\in V$ %  =\{0,1,..\} 
%random numbers $a_0,a_1,\dots$ 
is defined as 
\[
\theta(\rho) =\sum_{v\in V}% {n=0}^\infty 
p(v)\rho^{v},
\] where $p(v)$ is the probability of $v$, $\sum_{v \in V} p(v)=1$ and $\rho$ is a formal variable.

%\bor{
Let  $f(s)=\sum f_n s^{n}$ and $g(s)=\sum g_n s^{n}$ be generating functions, then {\em composition of generating functions} $f(s)$ and $g(s)$ is defined as $\sum_{n=0}^{\infty} f_{n}(g(s))^{n}$.  

\begin{lemma} \label{Lemma}
Let a non-random sequence $\bs u=(u_0,u_1,...)$, \irina{$u_i\ge 0$} 
have CGF $\psi(s)$ and  $\bs v=(v_{0},v_{1},...)$ be a sequence of \irina{independent identically distributed} discrete random  variables  $v_i\in V$ having  MGF $\theta(\rho)$ each. 
Introduce a new random variable  
\begin{equation} \label{L1}
z_j=\sum_{i=0}^{u_j}  v_i\;, \quad j=0,1,...   
\end{equation}
Then the average over all possible $\bs v$ CGF $\Psi(\rho)$ of the sequence $\bs z=(z_0,z_1,...)$ is equal to 
\begin{equation} \label{L2}
\Psi(\rho)=\psi(s)\vert_{s=\theta(\rho)}=\psi(\theta(\rho))\;.
\end{equation}
\end{lemma}
\begin{IEEEproof}
This lemma is a straightforward generalization of  formula (1.3) \cite[Chapter XII]{feller1968probability}.
\end{IEEEproof}

\subsection{Examples of SEDS for BICM,  BPCM and ASCM mapping \label{Apspectra}} 
\irina{

Examples of MGF of normalized SEDS $\alpha_{n,d}$ computed by using (\ref{alpha}) 
are given in Table  \ref{collection}. 
Each $\alpha_{n,d}$ is presented in the form of two tables, where the $i$-th row represents  the SEDs corresponding to  Hamming weight $i$ of the binary representation for a signal point. 
MGF is represented by a list of nonzero probabilities and  the corresponding normalized SEDs.  
For large signal sets only six  first coefficients of MGFs are presented.

Computations show that BPCM and ASCM for 8-PAM and $q=2^4$ are identical. 
Nevertheless, for practical codes simulation results differ. 
Notice that ASCM implies some restrictions on combination of modulation order and 
size of symbol alphabet, whereas BPCM is applicable for an arbitrary set of parameters. 
}

\begin{table} 
\centering
\caption{ MGF of normalized SEDS for selected mappings 
\label{collection}}
\arraycolsep=2pt\def\arraystretch{1.5}
\begin{tabular}{|c|c|c|c|c|}   \hline
Mapping &  \multicolumn{2}{c|}{Parameters }  &  Probabilities & Distances \\ \cline{2-3}
             &                               $q$   & $M$     &                      &          \\ \hline
 BICM     &    --                               &  4         &     
% $  \begin{pmatrix} 3/4&1/4 \\ 1&0 \end{pmatrix} $ &
 $  \begin{array}{rr} 3/4&1/4 \\ 1&0 \end{array} $ &
 $  \begin{array}{rr} 4&36 \\ 8& -\end{array} $  \\ \hline
BICM     &    --                               &  8        &     
 $  \begin{array}{rrrr} 7/12&1/4 &1/12&1/12 \\ 1&1/2&1/3& 1/6\\ 1/2 &1/2&- &- \end{array} $ &
 $  \begin{array}{rrrr} 4&36&100&196 \\ 8&32 & 72& -\\12&100/3&-&-\end{array} $  \\ \hline
SICM     &   4                               &  4         &     
 $  \begin{array}{rrrr} 3/4&1/4 &0&0 \\ 3/8 &1/3&1/4& 1/24\\ 3/4 &1/4&- &- \\1&-&-&-\end{array} $ &
 $  \begin{array}{rrrr} 4&36&-&- \\ 4&8 & 10& 18\\20/3&52/3&-&-\\8&-&-&-\end{array} $  \\ \hline
SICM     &   8                               &  6         &   \scriptsize{
\arraycolsep=1.4pt\def\arraystretch{1}
% $  \begin{pmatrix} 
 $\begin{array}{rrrrrrr}
 0.583 &     0.25  &  0.083 &   0.083 &        0  &       0 &\dots\\
0.207  &   0.203  &   0.177  &   0.135   &  0.038  &  0.0591&\dots\\
0.276  &  0.053 &    0.118 &    0.184  &   0.132 &   0.0395&\dots\\
0.158  &   0.123  &  0.053 &    0.211  &   0.123 &   0.0702&\dots\\
 0.25   &  0.167   &   0.25   &  0.167    & 0.083  &  0.0833&\dots\\
 0.25   &    0.5    &  0.25     &    0        &     0      &   0&\dots
 %\end{pmatrix} $
 \end{array} $
 } &  \scriptsize{
 \arraycolsep=1.4pt\def\arraystretch{1}
 %$  \begin{pmatrix} 
 $\begin{array}{rrrrrrr}
     4&    36&   100 &  196 &    - &    -1&\dots\\
     4 &   8&    20  &  32 &   36 &  52&\dots\\
    20/3 &   12&    52/3 &   68/3 &  100/3  & 116/3&\dots\\
    8 &   10&    18  &  20 &  26 &  32&\dots\\
    52/5 &  20&   116/5&   164/5&   36 &  244/5&\dots\\
    12 &  78/3&   100/3&    -&   -&    -&\dots
 %\end{pmatrix} $
 \end{array}$
 }  \\ \hline
\parbox{1cm}{BPCM, ASCM}     &   4                               &  8         &   \scriptsize{
\arraycolsep=1.4pt\def\arraystretch{1}
% $  \begin{pmatrix} 
$\begin{array}{rrrrrrr} 
0.583  &    0.25 &   0.083  &    0.083&         0&         0&\dots\\
 0.283 &   0.093 &    0.243 &   0.062&     0.052&    0.081&\dots\\
 0.138 &    0.203  &  0.013&     0.178&     0.087&     0.135&\dots\\
0.034&     0.202 &   0.049&    0.038&    0.059&    0.173&\dots\\
  0.11  &   0.162  &  0.063&    0.031&     0.141&    0.069&\dots\\
 0.113 &   0.066 &   0.086&    0.101&    0.038&     0.187&\dots\\
 0.106 &    0.124&    0.061&    0.024&    0.046&     0.106&\dots\\
0.033 &    0.151&    0.029&   0.029&    0.067&    0.112&\dots\\
0.093 &    0.109 &   0.007&    0.047&     0.186&    0.072&\dots\\
0.094 &   0.024&    0.011&     0.125 &    0.188 &   0.073&\dots\\
0.063&   0.042&     0.188 &    0.125 &    0.208 &    0.125&\dots\\
0.063&     0.25&     0.375 &     0.25 &   0.063&        0&\dots
\end{array}$
% \end{pmatrix} 
 } &  \scriptsize{
 \arraycolsep=1.4pt\def\arraystretch{1}
% $  \begin{pmatrix} 
$\begin{array}{rrrrrrr}
      4  &  36  & 100  & 196 & - & - & \dots\\
      4 &  8&    20&    32&    36&   52&\dots\\
     4&    20/3&    12&    44/3  &  52/3&    68/3&\dots\\
     4&    6 &    8&    10 &   12&    14&\dots\\
    28/5&    36/5&    44/5&    52/5&    12&    68/5&\dots\\
    20/3&    6&   28/3&    12&    40/3&    44/3&\dots\\
    52/7&    60/7&    68/7&    76/7&    14&    92/7&\dots\\
    8&    9&    10&    11&   12&   14&\dots\\
    84/9&    92/9&   12&   124/9&   44/3  &  140/9&\dots\\
   10.4 &  11.2 &  14.4&   15.2&   16.8&   17.6&\dots\\
   \frac{124}{11}&   \frac{172}{11}&   \frac{188}{11}&   \frac{236}{11}&   \frac{252}{11}&   \frac{300}{11}&\dots\\
   12&   52/3&    68/3 &   28&  100/3 &    -&\dots
\end{array}$   
 %\end{pmatrix} $
 }  \\ \hline
   \end{tabular}
\end{table}

%
%Consider an NB LDPC code over GF$(2^4)$ used with  8-PAM signaling. Assume that BPCM mapping is applied. Then, the smallest $n_{\rm s}$ and $n_p$ satisfying (\ref{compat}) are $n_{\rm s}=3$ and $n_p=4$. We measure the Hamming distance between the $\binom{mp_p}{2}$ pairs of  binary vectors $\vec v$ of length $m\times n_p=12$ and the  squared Euclidean distance between the corresponding  pairs of vectors $\vec s_{i}=(s_{i,1},s_{i,2},s_{i,3})$, $s_{i,j} \in \{-7,...,-1,1,...,7\}$,$j=1,2,3$.  The first ten terms of the generating function $g^{C}(\lambda)$ are given below
%
%\[g^{C}(\lambda)=0.175348\lambda^{4}+ 0.17202\lambda^{ 6}+0.0982971\lambda^{6.67}+0.0608506 \lambda^{8}+0.0327657\lambda^{10}+0.0564136\lambda^{12}\]
%\[+0.147446\lambda^{14}
%+ 0.08601\lambda^{14.67}+0.0421273\lambda^{17.33}+ 0.128722\lambda^{18}+...\]
%Similarly, for ASCM mapping used for an NB LDPC code over GF$(2^4)$ with 32-PAM signaling, we obtain 
%\[g^{D}(\lambda)=0.181765 \lambda^4+ 0.1791\lambda^6+ 0.0682285\lambda^{6.67}+ 0.0389877\lambda^8+    0.0227428 \lambda^{10}+  0.107724 \lambda^{12}\]
%\[+   0.153514\lambda^{14}+ 0.08955\lambda^{14.67}+0.0292408 \lambda^{17.33}+0.129147\lambda^{18}+...
%\]
%}
 \end{appendix}
 
\bibliographystyle{IEEEtran}
%\bibliography{IEEEabrv,refer}  
%\bibliographystyle{splncs04}
\bibliography{refer2}
\end{document}